\documentclass[12pt,preprint]{aastex}
\usepackage{color,graphicx}
\newcommand{\ie}{{\em i.e.}}

\newcommand{\newtext}{}
\begin{document}
\title{Initial Results from Fitting Resolved Modes using HMI Intensity
  Observations}
\author{S.~G.~Korzennik}
\affil{Harvard-Smithsonian Center for Astrophysics, Cambridge, MA 02138, USA}
\maketitle
\begin{abstract}
The HMI project recently started processing the continuum intensity images
following global helioseismology procedures similar to those used to process
the velocity images.  The spatial decomposition of these images has produced
time series of spherical harmonic coefficients for degrees up to $\ell=300$,
using a different apodization than the one used for velocity observations. The
first 360 days of observations were processed and made available. In this
paper I present initial results from fitting these time series using my state
of the art fitting methodology and compare the derived mode characteristics to
those estimated using co-eval velocity observations.
\end{abstract}

\section{Introduction}

Recently, the Helioseismic and Magnetic Imager (HMI) project started
processing the HMI continuum intensity images following procedures similar to
those used to process the MDI and HMI velocity images. This generated time
series of spherical harmonic coefficients suited for global helioseismology
mode fitting.

The spatial decomposition of apodized intensity images was carried out for the
first 360 days of the HMI science-quality data, producing time series of spherical
harmonic coefficients for degrees up to $\ell=300$. Since the oscillatory
signal in intensity is not attenuated by a line of sight projection, the
intensity images were apodized differently from the velocity images. Moreover,
since the global helioseismology data processing pipeline was developed using
velocity images, the automatic detection of discontinuities in the intensity
data has yet to be implemented and validated. For that reason the HMI project
has not yet applied its gap filling to the resulting time series.

{
While solar p-mode oscillations were detected in intensity decades ago by
\citet{Woodard:1984} with the ACRIM instrument on board SMM, the intensity
images in most spatially resolved experiments are not routinely
analyzed. Indeed, neither GONG, nor MDI or HMI pipelines process the intensity
images.

Historically, solar oscillation data have been acquired and analyzed using
intensity fluctuations for integrated observations 
\cite[see][for example]{2013ASPC..478..145S}. 
For a few cases, intensity images have been
reduced \citep{2013ASPC..478..151C} and in most cases a cross-spectral
analysis was carried out on $m$-averaged spectra and without the inclusion of
any spatial leakage information \citep{OlivieroEtal:2001,2004ApJ...602..516B}.

None of these studies led to a routine reduction and analysis of the intensity
images, since the ``noise'' properties of the intensity data are quite
different from the velocity data and fewer modes can be fitted. Nevertheless,
fitting intensity data allows for an independent validation of the fitting
methodology and further confirmation for the need to fit an asymmetric profile.
Indeed, the GONG, MDI and HMI pipelines are still fitting symmetric profiles
to mode peaks that are known to be asymmetric. Moreover the GONG pipeline
simply ignores the leakage matrix, while the MDI and HMI pipeline includes the
leakage matrix but continues to routinely fit symmetric profiles. 

The MDI and HMI mode fitting procedure was retrofitted to include an
asymmetry, but when using asymmetric profiles it fits fewer modes
successfully and it produces a more inconsistent set of modes with fitted
epoch.
Finally, the mode asymmetry measured by the MDI and HMI fitting procedure
barely changes with time or activity level, while the mode asymmetry
measured by my methodology shows changes that correlate with the solar activity
levels 
\cite[see][]{Korzennik:2013}.

By fitting the intensity and the velocity independently we can validate both
the inclusion of the leakage matrix and the proper modeling of the
asymmetry. Indeed, the intensity leakage is substantially different from the
velocity leakage and the mode frequency ought to be the same whether the
oscillatory signal is observed and measured in intensity or velocity. By
contrast a cross-spectral analysis {\newtext models both the intensity and the
  velocity spectra but fits a single parameter for the mode frequency, hence
  the velocity and intensity frequency is the same by construct.}

In this paper I present my first attempt to fit these time series, using my
state of the art fitting methodology
\citep{Korzennik:2005,Korzennik:2008}. While that method is in principle
perfectly suited to velocity or intensity observations, a leakage matrix
specific to intensity observations was needed.

I fitted {four} consecutive 72-day long time series of intensity observations as
well as one 288-day long time series (\ie, one four times longer). I carried out
my mode fitting using the same procedures as I use for velocity observations,
although I first refined the initial guess used for the mode profile asymmetry
to be appropriate for intensity observations, and used a leakage matrix
appropriate for intensity observations. I also ran my fitting procedure by
forcing the mode profile to be symmetric. Finally, in order to extend the
comparison to the 288-day long time series, I ran my fitting procedure on the
same co-eval 288-day long time series using symmetric mode profiles and
velocity observations.

I describe in Section~\ref{sec:descILM} the various leakage matrix coefficient
estimates I computed and/or used, and how I tried to validate them against the
observed power distribution with $m$.  The results from fitting intensity
observations are presented in Section~\ref{sec:resInt}, and I first compare,
in Section~\ref{sec:compILM}, the results obtained from fitting the same
intensity observations time series using two different leakage
matrices. Section~\ref{sec:compInt} shows comparisons between mode parameters
derived from fitting intensity and velocity observations, all using my fitting
methods, but also cases run leaving the mode profile symmetric.

\subsection{Data Set Used}

The data set used for this study are time series of spherical harmonic
coefficients computed by the HMI project at Stanford using the continuum
intensity images taken by HMI on board the Solar Dynamic Observer (SDO). This
data set is tagged at the SDO HMI and AIA Joint Science Operations Center
(JSOC) as {\tt hmi.Ic\_sht\_72d}. {Four} consecutive time series,
each 72-day long, were produced for degrees up to $\ell=300$ and for all
azimuthal orders, $m$, starting on 2010.04.30 at 00:00:00 TAI. These time
series were not gap-filled, although the fill factors are high, namely between
{97.078} and 99.660\%. One 288-day long time series was constructed using, for
consistency with previous analysis, {four 72-day long time series
  starting on 2010.07.11 TAI}
(\ie, 72$\times$72 days after the
start of the Michelson Doppler Imager, or MDI, science-quality data).
{
The start and end time of the fitted time series and their respective duty
cycles are listed in Table~\ref{tab:fitranges}.
}

\subsection{Brief Description of the Fitting Methodology}
\label{sec:BDFM}

My state of the art fitting methodology is described at length in
\cite{Korzennik:2005,Korzennik:2008}. The first step consists in computing
sine multi-taper power spectra, with the number of tapers optimized to match
the anticipated effective line-width of the modes being fitted, 
{
hence the number of tapers is not constant for a given time series 
length\footnote{
 For 72-day long time series, the number of tapers is between
 3 and 33 (i.e., 3, 5, 9, 17 or 33) while for the 288-day long time series it
 is between 3 and 129 (i.e., 3, 5, 9, 17, 33, 65 or 129).}
\citep[see][for details]{Korzennik:2005}.}
The second
step consists in fitting simultaneously all the azimuthal orders for a given
mode, using a fraction of the power spectrum centered around the fitted
mode. Each singlet, \ie: ($n,\ell,m$), is modeled by an asymmetric mode
profile characterized by its own frequency, amplitude and background, and by a
line-width and asymmetry that is the same for all azimuthal orders, 
{hence fitted model assumes that the FWHM and the asymmetry are independent of $m$}
The fitted model includes a complete leakage matrix, where the leaked modes,
modes for the same $n$ but a different $\ell$ and $m$, are attenuated by the
ratio of the respective leakage matrix components. Contamination by nearby
modes, namely modes with a different $n$, $\ell$ and $m$, is also included in
the model when these modes are present in the spectral fitting window.

The model is fitted simultaneously, in the least-squares sense, to the observed
$2\ell+1$ multi-tapered power spectra. For numerical stability the fitting is
done in stages, \ie, not all the parameters are fitting simultaneously right
away, and a sanity check is performed along the way: modes whose amplitude is
not above some threshold based on the spectrum SNR are no longer fitted.  A
third step consists in iterating the fitting of each mode using the results of
the previous iteration to account for the mode contamination.

Sections of power spectra, $P_{n,\ell,m}(\nu)$ are modeled as
\begin{eqnarray}
P_{n,\ell,m}(\nu) & = & \Sigma_{\ell',m'} \left(
    \frac{C(\ell,m;\ell',m')}{C(\ell,m;\ell,m)} 
      A_{n,\ell',m'} {\cal L}(\frac{\nu-\nu_{n,\ell',m'}}{2\,\Gamma_{n,\ell'}}, \alpha_{n,\ell''}) +
      B_{n,\ell',m'} \right) \\ 
& & + \Sigma_{n'} P_{n',\ell,m}(\nu)
\end{eqnarray}
where $\nu$ is the frequency, ${\cal L}$ a generalized asymmetric Lorentzian,
defined as
\begin{equation}
{\cal L}(x, \alpha) = \frac{1 + \alpha ({x}-\frac{\alpha}{2})}{1+{x}^2}
\end{equation}
and $\nu_{n,\ell,m}, \Gamma_{n,\ell}, \alpha_{n,\ell}, A_{n,\ell,m}$, and
$B_{n,\ell,m}$ are the mode frequency, FWHM, asymmetry, power amplitude, and
background respectively, while $C(\ell,m;\ell',m')$ are the leakage matrix
coefficients.

\subsection{Intensity Leakage Matrix}
\label{sec:descILM}

\subsubsection{Sensitivity Function and Limb Darkening}
By contrast to the velocity oscillatory signal 
\citep[see, for example,][]{Korzennik:2005}, the
intensity oscillatory signal is a scalar, leading to a simpler leakage matrix,
namely:
\begin{equation}
C(\ell, m; \ell',m') = \int {\cal A}(\mu) J(\mu)\,
    Y_{\ell}^{m*}(\theta,\phi)\,Y_{\ell'}^{m'}(\theta,\phi)\, d\Omega
\end{equation}
where $\theta$ is the co-latitude, $\phi$ the longitude, $\mu$ the fractional
radius of the image of the solar disk, $\cal A$ the apodization used in the
spatial decomposition, $J$ the sensitivity of the oscillatory signal,
$Y_{\ell}^{m}$ the spherical harmonic of degree $\ell$ and azimuthal order
$m$, and $d\Omega= \sin\theta d\theta d\phi$. The integral extends in $\theta$
and $\phi$ to cover the visible fraction of the Sun.

The sensitivity function, $J$, is likely to be equivalent to the limb
darkening function, $I$, although this ought to be checked.  In principle, the
sensitivity function can be empirically computed from the observations by
computing the RMS of the oscillatory signal as a function of position on the
solar disk and reducing it to a function of $\mu$, the fractional radius.
Hence, I computed the RMS of the residual intensity signal, after detrending
the images, using HMI continuum images taken on ten consecutive days, for six
different years. I detrended the images using a 15-minute long running mean,
then computed, using the time series of residuals images, the mean and RMS
around the mean of the residual signal, rebinned as a function of fractional
radius, $\mu$, and normalized to unity at disk center.  The solar limb
darkening, for a set of wavelengths, has been measured and is reported in
\cite{Pierce+Slaughter:1977}.

The empirical sensitivity functions I derived for each year, the average for
the six years, and the limb-darkening profiles given in
\cite{Pierce+Slaughter:1977} interpolated at $\lambda=617.3$ nm, the
wavelength HMI is observing at
\citep{2012SoPh..275..229S,2012SoPh..275..285C,2016SoPh..291.1887C}, and the
profiles used by the Stanford group (private communication) are all compared
in Fig.~\ref{fig:limb}. One additional complication is the behavior near the
limb of the different formulations of the polynomial representation of the
limb-darkening, given either as a function of $x=\ln(\mu)$ or $\mu$; see Tables
II or IV of \cite{Pierce+Slaughter:1977}.

Since the intensity oscillatory signal is not attenuated by the line of sight
projection, the apodization for the intensity images could be pushed closer to
the edge of the solar disk without substantially adding noise, like in the
case of velocity. The apodization was chosen by the Stanford group to start at
$\mu=0.98$, consisting of a cosine bell attenuation that spans a range in
$\mu$ of $0.015$, as indicated by the vertical lines drawn in
Fig.~\ref{fig:limb}.


The different profiles shown in Fig.~\ref{fig:limb} are somewhat similar. Note
how the empirical sensitivity profiles resulting from processing each of the
six years are nearly identical. They deviate from the limb-darkening profiles,
suggesting an increased sensitivity for $0.3 \le \mu \le 0.6$, and a sharper
decrease in sensitivity for $\mu \ge 0.8$. In contrast, the different limb
darkening profiles are almost identical for $\mu < 0.9$, except that the
polynomial parametrization in $x=\ln(\mu)$ leads to negative values close to
the limb, including the one based on the Stanford version 2 coefficients. The
polynomial parametrization in $\mu$ of the limb-darkening does not include the
progressive attenuation near the limb resulting from an empirical
determination of the sensitivity profile, although the contribution to the
leakage matrix of the regions with $\mu \ge 0.98$ is dominated by the
apodization.

The precise profile to be used for the computation of the intensity
leakage matrix is yet to be determined. I opted to use a polynomial
parametrization in $\mu$, and either the limb-darkening, $I(\mu)$, given by
the coefficients in Table IV of \cite{Pierce+Slaughter:1977}, interpolated at
$\lambda=617.3$ nm, or a polynomial in $\mu$ fitted to my determination of the
averaged empirical sensitivity function, $\bar{J}(\mu)$, for all six processed
years. I also used the leakage matrix computed by the HMI group at Stanford
(Larsen, private communication).

\subsubsection{Computation and Validation of the Leakage Matrix}

A leakage matrix is ``{\em simply}'' computed by generating images representing
the quantity $J(\mu)\, Y_{\ell}^{m*}(\theta,\phi)$, or $I(\mu)
Y_{\ell}^{m}(\theta,\phi)$ and processing them using the same spatial
decomposition used for the observations.

The effects of the actual orientation, {\em i.e.}, $P_{\rm eff}$, the
effective position angle and $B_o$, the latitude at disk center,
$D^o_{\odot}$, the finite observer to Sun distance, and the image
pixelization, while not described explicitly here, are taken into account when
computing the images that are decomposed to generate a leakage matrix
{
\citep[see][]{KorzennikEtal:2004, Schou:1999}
}. My
computation evaluated $C_{\ell,m}(\delta\ell,\delta m) = C(\ell,m;\ell', m')$
for $\delta\ell=\pm20$ and $\delta m=\pm20$, while the HMI group at Stanford
limited their evaluation to $\delta\ell=\pm6$ and $\delta m=\pm15$, where
$\delta\ell=\ell'-\ell$ and $\delta m=m'-m$.

In an attempt to validate the different computations of leakage matrices
suited for intensity observations, I choose to compare the variation with
respect to $m$ (or the ratio ${m}/{\ell}$) of the leakage to the variation
of the observed power.

We can assume that the mode amplitude ought to be uniform with $m$, in the
absence of any physical mechanism that would modulate the amplitude with
$m$. If this is indeed the case, the variation of the observed total power, or
the measured power amplitude of the modes, is only the result of the variation
of the leakage matrix with $m$. Therefore the total power
variation with $m$ at a fixed $\ell$ should be proportional to the sum of
sensitivity of the target 
mode plus the contribution of the leaks.  We can thus equate the normalized
total power
\begin{equation}
  \bar{P}_{\ell,m}^{\rm Tot} = \frac{1}{P_{N}} = \Sigma_{\nu}\,  P_{\ell,m}(\nu)
\end{equation}
to
\begin{equation}
 \bar{Q}_{\ell,m}^{\rm Tot} = \frac{1}{Q_{N}} \Sigma_{\delta\ell,\delta m}\,C^2_{\ell,m}(\delta\ell,\delta m)
\end{equation}
where $P_{N}$ and $Q_{N}$ are normalization factors chosen to set
$\bar{Q}_{\ell,m=0}^{\rm Tot} = \bar{P}_{\ell,m=0}^{\rm Tot} = 1$. 
 
On the other hand, the modes observed power amplitude, $A_{n,\ell,m}$, as
measured by fitting the modes, should be proportional to the values of the
$\delta\ell=\delta m=0$ leak, or $C^2_{\ell,m}(0,0)$. Hence the quantity
\begin{equation}
  \bar{A}_{\ell,m} = \frac{1}{A_{N}} \Sigma_{n}\,  A_{n,\ell,m}
\end{equation} is equal to the ratio
\begin{equation}
  \bar{Q}_{\ell,m}=\frac{C^2_{\ell,m}(0,0)}{C^2_{\ell,m=0}(0,0)}
\end{equation} 
if $A_{N}$ is such that $\bar{A}_{\ell,m=0} = 1$, since $\bar{Q}_{\ell,m=0}=1$ by construction.

In order to build statistical significance for the observed quantities
$\bar{P}^{\rm Tot}_{\ell,m}$ and $\bar{A}_{\ell,m}$, I performed additional
averaging over a range in $\ell$ ($\delta\ell=\pm1$), plus some smoothing
over $m$ and symmetrization in $m$.

Figures~\ref{fig:valid-1a} to \ref{fig:valid-2} show these comparisons, using
three distinct leakage matrices and a set of degrees. While the
overall variation with $m/\ell$ agrees qualitatively, none of the leakage
matrices lead to $\bar{Q}^{\rm Tot}_{\ell,m}$ or $\bar{Q}_{\ell,m}$ profiles
that closely match the observed quantities, $\bar{P}^{\rm Tot}_{\ell,m}$ or
$\bar{A}_{\ell,m}$ respectively. Moreover, the two methods do not agree as to
which case models best the observed quantities. This apparent contradiction
could be the result of the wrong assumption that the mode power is independent
of $m$. Since it is the solar rotation that breaks the spherical symmetry and
thus ``defines'' $m$, it is not inconceivable that, while the solar rotation
is slow compared to the oscillations, 
the rotation attenuates some azimuthal orders over others and produces
an intrinsic variation of the modes amplitude with azimuthal order, $m$.


\subsection{Seed Asymmetry for Intensity}

Using high degree resolved modes, \cite{DuvallEtAl:1993} were the first to
notice that not only are the profiles of the modes asymmetric, but the
asymmetry for velocity observations is of the opposite sign than the asymmetry
for intensity observations. This asymmetry is, of course, also present at low
and intermediate degrees, and is expected to be of opposite sign for velocity
and intensity.

For each mode set, the fitting starts from some initial guess, also known as a
seed. The seed file holds the list of modes to attempt to fit, \ie, the
coverage in $(n,\ell)$, and for each mode a rather good initial guess of the
mode's central frequency, or multiplet, the frequency splitting parametrized by
a polynomial expansion in $m$, its line-width and its asymmetry. The initial
guesses for the asymmetry are set to be a smooth function of frequency, and
for velocity observations, using my parametrization, are mostly negative. Since
the asymmetry of the intensity observations is of the opposite sign, a new
seed asymmetry had to be computed.

To accomplish this, I ran my second step, or initial fit, as described earlier
in Section~\ref{sec:BDFM}, using one 72-day long segment, and using at first
the negative initial guesses for $\alpha$ appropriate for velocity
observations, \ie, $\alpha_{n,\ell}^{sV}$. The resulting fitted asymmetries
were mostly positives. I proceeded to fit a polynomial in $\nu$ to them and
produced an updated seed file with new initial guesses for intensity
observations, \ie, $\alpha_{n,\ell}^{sI}$. I repeated this procedure six
times, as illustrated in Fig.~\ref{fig:seed-alpha}, until the resulting mean
change in the resulting fitted frequencies was negligible. The final
parametrization of the initial guess for $\alpha_{n,\ell}^{sI}$ was
subsequently used to fit all the intensity observations.


\section{Fitting Results}
\label{sec:resInt}

For reasons of convenience explained earlier, the times series of spherical
harmonic coefficients computed by spatially decomposing HMI continuum
intensity images have not been gap filled. I computed sine multi-tapered power
spectra for {four} consecutive 72-day long time series and one 288-day long time
series. The power spectra were fitted using my fitting methodology, using the
seed file adjusted to take into account the mode profile asymmetry for
intensity observations, and two sets of leakage matrices: one computed by
myself based on the limb-darkening parametrized by a 5 coefficient polynomial
in $\mu$ 
\citep[][interpolated at $\lambda=617.3$ nm]{Pierce+Slaughter:1977}
and one provided by the HMI group at Stanford, courtesy of Drs.\ Larson and
Schou (private communication).

Only the 72-day long time series were fitted using both leakage matrices, and
using an asymmetric profile. All the other cases were fitted using only the
leakage matrix I computed, based on a limb-darkening profile. In order to
assess the effect of fitting the asymmetry, I also fitted the intensity data
with a symmetric profile. This was accomplished by modifying the seed file to
set the asymmetry to zero, and changing the steps used in the fitting
procedure to leave the asymmetry parameter null by never including it in the
list of parameters to fit.

\subsection{Intensity SNR Limitation}

A major difference between velocity and intensity oscillatory signals, besides
the sign of the asymmetry, is the nature of the so-called background noise, so
called because it is a signal of solar origin that adds a noisy background
level to the oscillatory signal. Intensity observations, whether disk
integrated or resolved,
show a noise contribution that increases as
the frequency decreases, of a $\nu^{-1}$ nature. The detrending that was
adequate for the velocity signal is no longer optimal for intensity, hence I
modified the detrending I perform on the time series before computing the sine
multi-taper power spectrum, from subtracting a 20-minute long running mean to
subtracting an 11-minute long running mean. This filters out power below 1.52
$\mu$Hz rather than below 0.83 $\mu$Hz.

Since my fitting methodology performs a sanity check at regular intervals,
modes at low frequencies, where the background level is high for intensity
observations, are no longer fitted. 
This attrition at low frequencies is illustrated in Fig.~\ref{fig:lnu}, where
the $(n,\ell,m)$ singlets that were successfully fitted are shown in a
$\ell-\nu$ diagram, and compared to the same representation when fitting a
similar data set derived from gap-filled velocity observations.


Because the coverage in the $\ell-\nu$ space is a lot more sparse for
intensity, I revised the procedure I use to derive multiplets, \ie,
$(n,\ell)$, from singlets. That procedure fits a Legendre polynomial to all
the successfully fitted frequencies, $\nu_{n,\ell,m}$, for a given $(n, \ell)$
mode as a function $m$ to derive a mode frequency, $\nu_{n,\ell}$, and
frequency splitting coefficients. The procedure fits from one to 9
coefficients, performs a 3-sigma rejection of outliers, and computes a mode
multiplet if and only if at least 1/8th of all the expected $m$ are used in
the polynomial. This criteria worked fine when fitting velocity observations,
but it eliminates most of the low-order, low-frequency modes, including all
the $f$-modes when fitting intensity observations.

I re-adjusted this procedure to derive a second set of multiplets using a less
stringent constraint, namely that at least {\em only} 1/16th of all the $m$
could be fitted. This led to some outliers that were then cleaned out by
eliminating modes whose frequency do not fall on a smooth function of $\ell$
for each order, $n$. This is illustrated in Fig.~\ref{fig:lnu} by the green
dots.

\subsection{Effect of Gap Filling and Longer Time Series on Low Frequency Noise}

Since the time series of intensity spherical harmonic coefficients were not
gap filled, I checked the contribution of the gaps to the background noise. A
naive estimate, illustrated in Fig.~\ref{fig:gap-noise}, suggests that gaps
scatter a lot of power into a higher background noise, including at low
frequencies. I therefore adapted the gap filler I use for the GONG
observations to gap fill {one 72 day long time series of}
HMI intensity data. This gap filler is the same
as the one used by the Stanford group to gap fill the MDI and HMI velocity
data. 


Figures~\ref{fig:gap-filled}, \ref{fig:show-spc-72d} and
\ref{fig:show-spc-288d} show that both gap filling and using longer time
series do not reduce the low frequency background noise. 
{
Fig.~\ref{fig:gap-filled} shows that (i) gap filling the intensity
observations barely changes the background levels; (ii) the background level
for intensity is about 20 times higher around 2 mHz than for velocity; and
(iii) the longer time series do not lower the background but reduce the
background realization noise. For the intensity observations, that reduction
is not sufficient to {see} the low-order, low-frequency modes. Note also
the clearly visible change of sign of the mode profiles asymmetry between
intensity and velocity power spectra.

Figures~\ref{fig:show-spc-72d} and \ref{fig:show-spc-288d} show (i) how the
realization noise produces spikes that without proper ``sanity check'' can
be easily confused as low amplitude modes, and (ii) that some modes peak above
the noise in an $m$-averaged spectrum but can't be discriminated from the
noise when fitting singlets.

From these figures, one concludes that the} %
power at low frequency is of solar origin and masks the oscillatory
signal. The power scatter by the gaps at these frequencies is negligible,
while increasing the length of the time series decreases the realization
noise, but not the background level. Eventually, a very long time series may
bring the realization noise to a level low enough to see a weak oscillatory
signal emerge clearly above the background, but quadrupling the length is not
enough. In fact, and somewhat counter-intuitively, quadrupling the length of
the time series resulted in making fitting low frequency modes more difficult.

{ For completeness, I also fitted the 288-day long time series using
  gap-filled time series. As anticipated, the resulting number of fitted modes
  and their characteristics are barely different from the raw data: a few more
  singlets were fitted but the same number of multiplets were derived when the
  observations are gap filled. The mean of the difference between raw and
  gap-filled data in the derived frequencies is less than 1 nHz, with a
  standard deviation of 13 nHz and differences in the derived FWHM and
  asymmetry are negligible.  }


\subsection{Results from 72-day and 288-day long Fitting}

Figures~\ref{fig:results-72d} and \ref{fig:results-288d} show mode
characteristics resulting from fitting 72-day and 288-day long
time series, after converting singlets to multiplets.
{
Table~\ref{tab:fitcstats} lists the number of fitted modes (singlets) and the
number of derived multiplets for each fitted time series, the different type of
data and leakage matrix used.
}
The FWHM, $\Gamma_{n,\ell}$, asymmetry, $\alpha_{n,\ell}$, the uncertainty of
the fitted frequencies, $\sigma_{\nu_{n,\ell}}$ and the mode power amplitudes,
$\bar{A}_{n,\ell}$, are plotted for the resulting multiplets, for one
{representative} 72-day long set and for the 288-day long set. The
corresponding values derived from fitting co-eval velocity observations are
shown as well.

Except for the low-order low-frequency modes, the FWHM and the frequency
uncertainties derived using either velocity or intensity observations agree
quite well. As expected, the asymmetry derived from intensity observations is
of opposite sign of the asymmetry derived from velocity observations but it is
also larger in magnitude by about a factor two.  The mode power amplitude
variation with frequency is overall similar, whether measured using intensity
or velocity observations, as it peaks at the same frequency but shows a
somewhat different distribution. This is most marked for results from fitting
72-day long time series and at low frequencies. Most of the extra
low-frequency modes derived from the 72-day long time series, using a less
stringent constraint to derive multiplets, show consistent values that mostly
agree with their velocity counterparts, except for higher uncertainties and
larger FWHM at the lowest frequencies. The higher uncertainty in itself is not
surprising since these multiplets are derived from fewer singlets, but the
increase in FWHM cannot easily be explained.

Contrasting results from fitting 72-day long time series to those resulting
from fitting 288-day long ones leads to the following observations: the mode
FWHM, frequency uncertainty, asymmetry and power amplitude distribution are
comparable, although (1) very few low frequency modes are successfully
derived; (2) the frequency uncertainty is reduced as expected by about a
factor 2, namely the square root of the ratio of the time series lengths; and
(3) the scatter in the measured asymmetry is reduced for intensity as it is
for velocity.

I have yet to fully understand why, when using the longer time series, almost
no modes below $\nu < 1800$ $\mu$Hz or $\Gamma < 0.8$ $\mu$Hz could be fitted
(see Fig.~\ref{fig:lnu}). This may suggest that despite appearing consistent,
the low frequency modes derived using a shorter time series are suspicious and
the methodology, especially the sanity check, needs to be adapted to the
specifics of the noise distribution of the intensity signal.


\subsection{Comparison using Different Leakage Matrices}
\label{sec:compILM}

Figures~\ref{fig:compare-leakage-72d} and \ref{fig:compare-leakage-288d} shows
a comparison of the mode parameters inferred by fitting the same time series
of intensity observations, using the exact same methodology but two different
estimates of the leakage matrix. Despite the different signature of the
leakage sensitivity with $m$, the resulting fitted frequencies, and most of
the other modes parameters, are barely different and show no systematic
trends.  Comparisons of the singlets frequency, or the singlets
scaled\footnote{The scaling is done by dividing the difference by its
  uncertainty.} frequency show a normal distribution with no significant bias
and a very low scatter. Only the mode line-width, $\Gamma$, when fitting the
longer time series, is systematically different, although not significantly.  Of
course, we cannot rule out that fitting much longer time series may lead to
small but significant or systematic differences. Still, this comparison shows
that for 72 and for 288-day long time series, the use of different leakage
matrix estimates does not really affect the fitted values.


\subsection{Comparison with Results from Fitting Velocity}
\label{sec:compInt}

Now that we have, for the first time, mode parameters resulting from fitting
the same interval based on either velocity or intensity HMI observations, let
us compare in detail the resulting mode characteristics.  Despite the fact
that the velocity time series were gap filled, while the intensity ones were
not, we have shown that we can rule out that this affected the results and
thus this comparison, because (i) the fill factors are already high; and (ii)
the background signal at low frequency is any way much higher for intensity
than for velocity.

Figures~\ref{fig:compare-vel-int-72d} and \ref{fig:compare-vel-int-288d}
compare frequencies, scaled frequencies, scaled FWHM and scaled asymmetries
derived from co-eval time series from either intensity or velocity
observations, for singlets or multiplets. The frequency comparisons show
virtually no bias for the singlets, but some small bias for the multiplets
({\em i.e.}, $0.43$ and $0.86\,\sigma$ for 72-day and 288-day long time series
respectively). Of course, the asymmetry differences are large and show a
smooth trend with frequency.


Since I also fitted the data using a symmetric mode profile, I can do the
exact same comparison but using mode characteristics derived from fitting a
symmetric profile for either type of observations or length of time
series. This comparison is presented in Figs~\ref{fig:compare-vel-int-sym-72d}
and \ref{fig:compare-vel-int-sym-288d} and systematic differences with skewed
distributions are clearly visible.

Table~\ref{tab:diffs} summarizes the comparisons and lists the mean and
standard deviation around the mean of the differences or scaled
differences. Comparing results from fitting symmetric profiles demonstrate
clearly the need to include the asymmetry of the mode profile at low and
intermediate degrees, and not just at high degrees. While the differences are
not very large in themselves, especially for 72-day long times series
singlets, they rise to the $2.3$ and $5.9\,\sigma$ levels for multiplets
derived from 72-day and 288-day long time series respectively, but more to the
point these differences clearly show systematic trends.
{ Close scrutiny of the table indicates a small residual bias in
  frequency differences from fitting co-eval velocity and intensity, even when
  using an asymmetric profile. It may well be that this small bias results
  from some remaining inadequacy in the fitting methodologies worth
  pursuing. This should not distract from the main conclusion that the
  inclusion of the asymmetry is key in the determination of accurate mode
  characteristics that are consistent whether measured using their
  manifestation from intensity or velocity fluctuations.}

\section{Conclusions}

Initial results from fitting HMI intensity observations using my state of the
art fitting methodology and including the mode profile asymmetry show a
remarkable agreement of the derived mode characteristics with the
corresponding values derived from co-eval velocity observations. Of course,
the mode asymmetry for intensity is of opposite sign to the the mode asymmetry
for velocity, as anticipated, and it is also larger in magnitude. The
comparison of mode frequency and FWHM determinations based on intensity and
velocity show no bias with a uniform normal distribution with a $0.3\,\sigma$
spread, and a very similar precision on the mode frequency. This being said,
my attempt to validate various estimates of the leakage matrix for intensity
shows residual inconsistencies that need to be resolved. I also show that
despite these inconsistencies, the derived modes characteristics do not seem to
be affected in any systematic way, at least for the precision resulting from
fitting 72-day or 288-day long time series. Fitting a much longer time series
may point to systematic errors associated to the leakage matrix determination.

One of the main drawbacks of intensity observations is the much higher noise
level at low frequencies than in velocity observations. For reasons that I
have yet to understand, and thus warrant more work, my fitting methodology was
able to determine low-order low-frequency singlets for the shorter time
series, but not for the longer one. One simple explanation could be that the
sanity rejection is not stringent enough and the fitted modes are just
realization noise spikes that happened to coincide with a mode frequency and
should be ignored. The principle that I have followed, namely to fit time
series of different lengths, again proves to be a good idea. I expect to fit
additional HMI intensity data as they become available and fit them using the
factor progression I have used for the velocity observations, namely fitting
time series that are 36-day, 72-day, 144-day, 288-day, etc... long.

Finally, comparisons of mode characteristics derived by fitting a symmetric
mode profile show unequivocally the systematic bias introduced in the mode
frequency determinations by ignoring the asymmetry. Also, by fitting
additional HMI intensity observations that will cover most of Cycle 24, I will
be able to confirm whether the mode asymmetry both for intensity and velocity
changes with solar activity, changes that I see in my fitting of velocity
observations, but is not seen by others. Indeed, co-eval intensity and
velocity derived frequencies ought to agree consistently independently of the
solar activity level. Therefore, a change in the velocity-derived asymmetry
will have to be matched by a change in the intensity-derived asymmetry,
although of opposite sign and different in magnitude, to keep the derived
frequencies in agreement.

\acknowledgements 
HMI data courtesy of NASA and the HMI consortium; HMI is supported by NASA
contract NAS5--02139 to Stanford University. The author wishes to thank
Drs.\ Larson and Schou for providing their estimate of the intensity leakage
matrix.  Dr.\ Korzennik is supported by NASA grant NNX15AL65G.
%

\begin{thebibliography}{15}
\expandafter\ifx\csname natexlab\endcsname\relax\def\natexlab#1{#1}\fi

\bibitem[{{Barban} {et~al.}(2004){Barban}, {Hill}, \&
  {Kras}}]{2004ApJ...602..516B}
{Barban}, C., {Hill}, F., \& {Kras}, S. 2004, \apj, 602, 516

\bibitem[{{Corbard} {et~al.}(2013){Corbard}, {Salabert}, {Boumier}, \& {Picard
  Team}}]{2013ASPC..478..151C}
{Corbard}, T., {Salabert}, D., {Boumier}, P., \& {Picard Team}. 2013, in
  Astronomical Society of the Pacific Conference Series, Vol. 478, Fifty Years
  of Seismology of the Sun and Stars, ed. K.~{Jain}, S.~C. {Tripathy},
  F.~{Hill}, J.~W. {Leibacher}, \& A.~A. {Pevtsov}, 151

\bibitem[{{Couvidat} {et~al.}(2016){Couvidat}, {Schou}, {Hoeksema}, {Bogart},
  {Bush}, {Duvall}, {Liu}, {Norton}, \& {Scherrer}}]{2016SoPh..291.1887C}
{Couvidat}, S., {Schou}, J., {Hoeksema}, J.~T., {Bogart}, R.~S., {Bush}, R.~I.,
  {Duvall}, T.~L., {Liu}, Y., {Norton}, A.~A., \& {Scherrer}, P.~H. 2016,
  \solphys, 291, 1887

\bibitem[{{Couvidat} {et~al.}(2012){Couvidat}, {Schou}, {Shine}, {Bush},
  {Miles}, {Scherrer}, \& {Rairden}}]{2012SoPh..275..285C}
{Couvidat}, S., {Schou}, J., {Shine}, R.~A., {Bush}, R.~I., {Miles}, J.~W.,
  {Scherrer}, P.~H., \& {Rairden}, R.~L. 2012, \solphys, 275, 285

\bibitem[{{Duvall} {et~al.}(1993){Duvall}, {Jefferies}, {Harvey}, {Osaki}, \&
  {Pomerantz}}]{DuvallEtAl:1993}
{Duvall}, Jr., T.~L., {Jefferies}, S.~M., {Harvey}, J.~W., {Osaki}, Y., \&
  {Pomerantz}, M.~A. 1993, \apj, 410, 829

\bibitem[{{Korzennik}(2005)}]{Korzennik:2005}
{Korzennik}, S.~G. 2005, \apj, 626, 585

\bibitem[{{Korzennik}(2008)}]{Korzennik:2008}
{Korzennik}, S.~G. 2008, in Journal of Physics Conference Series, Vol. 118,
  Journal of Physics Conference Series, 012082

\bibitem[{{Korzennik}(2013)}]{Korzennik:2013}
{Korzennik}, S.~G. 2013, in Astronomical Society of the Pacific Conference
  Series, Vol. 478, Fifty Years of Seismology of the Sun and Stars, ed.
  K.~{Jain}, S.~C. {Tripathy}, F.~{Hill}, J.~W. {Leibacher}, \& A.~A.
  {Pevtsov}, 137

\bibitem[{{Korzennik} {et~al.}(2004){Korzennik}, {Rabello-Soares}, \&
  {Schou}}]{KorzennikEtal:2004}
{Korzennik}, S.~G., {Rabello-Soares}, M.~C., \& {Schou}, J. 2004, \apj, 602,
  481

\bibitem[{{Oliviero} {et~al.}(2001){Oliviero}, {Severino}, \&
  {Straus}}]{OlivieroEtal:2001}
{Oliviero}, M., {Severino}, G., \& {Straus}, T. 2001, in ESA Special
  Publication, Vol. 464, SOHO 10/GONG 2000 Workshop: Helio- and
  Asteroseismology at the Dawn of the Millennium, ed. A.~{Wilson} \& P.~L.
  {Pall{\'e}}, 669--672

\bibitem[{{Pierce} \& {Slaughter}(1977)}]{Pierce+Slaughter:1977}
{Pierce}, A.~K. \& {Slaughter}, C.~D. 1977, \solphys, 51, 25

\bibitem[{{Salabert} {et~al.}(2013){Salabert}, {Garc{\'{\i}}a}, \&
  {Jim{\'e}nez}}]{2013ASPC..478..145S}
{Salabert}, D., {Garc{\'{\i}}a}, R.~A., \& {Jim{\'e}nez}, A. 2013, in
  Astronomical Society of the Pacific Conference Series, Vol. 478, Fifty Years
  of Seismology of the Sun and Stars, ed. K.~{Jain}, S.~C. {Tripathy},
  F.~{Hill}, J.~W. {Leibacher}, \& A.~A. {Pevtsov}, 145

\bibitem[{{Schou}(1999)}]{Schou:1999}
{Schou}, J. 1999, \apjl, 523, L181

\bibitem[{{Schou} {et~al.}(2012){Schou}, {Scherrer}, {Bush}, {Wachter},
  {Couvidat}, {Rabello-Soares}, {Bogart}, {Hoeksema}, {Liu}, {Duvall}, {Akin},
  {Allard}, {Miles}, {Rairden}, {Shine}, {Tarbell}, {Title}, {Wolfson},
  {Elmore}, {Norton}, \& {Tomczyk}}]{2012SoPh..275..229S}
{Schou}, J., {Scherrer}, P.~H., {Bush}, R.~I., {Wachter}, R., {Couvidat}, S.,
  {Rabello-Soares}, M.~C., {Bogart}, R.~S., {Hoeksema}, J.~T., {Liu}, Y.,
  {Duvall}, T.~L., {Akin}, D.~J., {Allard}, B.~A., {Miles}, J.~W., {Rairden},
  R., {Shine}, R.~A., {Tarbell}, T.~D., {Title}, A.~M., {Wolfson}, C.~J.,
  {Elmore}, D.~F., {Norton}, A.~A., \& {Tomczyk}, S. 2012, \solphys, 275, 229

\bibitem[{{Woodard}(1984)}]{Woodard:1984}
{Woodard}, M.~F. 1984, PhD thesis, University of California, San Diego.

\end{thebibliography}

%
\newcommand{\Clearpage}{\clearpage}

\Clearpage
\begin{figure}
\begin{center}
\includegraphics[width=.975\textwidth]{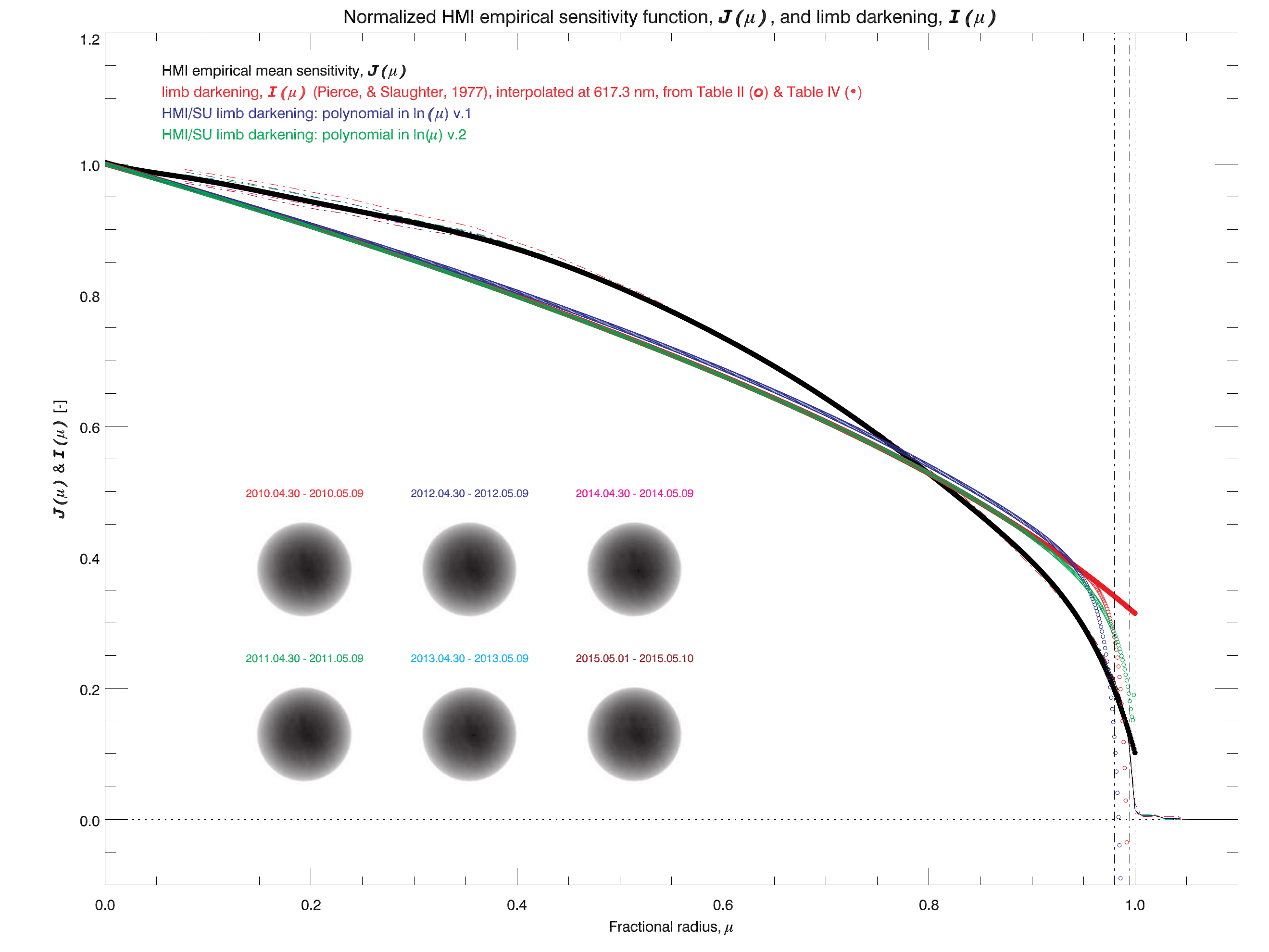}
\end{center}
\caption{Empirical sensitivity functions, $J(\mu)$, and limb-darkening
  functions, $I(\mu)$, as a function of the fractional radius, $\mu$. The
  colored dash curves are estimates of $J(\mu)$ derived from 10 days of data
  taken in six consecutive years. The images of the RMS of the residuals used
  for this derivation are shown with the corresponding color-coded time
  ranges. The black solid line is the average of these six profiles, and the
  black dots the corresponding polynomial fit to this average that was used
  for one leakage matrix computation.
  The colored dots and circles correspond to limb-darkening profiles computed
  using different polynomial parametrization: Tables II and IV from
  \cite{Pierce+Slaughter:1977}, interpolated for $\lambda=617.3$ nm, and
  coefficients used by the Stanford group (private communication). Open
  circles correspond to polynomials in $x=\ln(\mu)$, dots to polynomials in
  $\mu$. Note how the different limb-darkening representations disagree only
  near the limb, and that the polynomial parametrization with respect to
  $x=\ln(\mu)$ leads to negative values near the limb.
  Vertical lines are drawn to indicate the location of the limb and the edges
  of the cosine bell apodization used for the intensity observations.
\label{fig:limb}}
\end{figure}

\Clearpage
\begin{figure}
\begin{center}
\includegraphics[width=.975\textwidth]{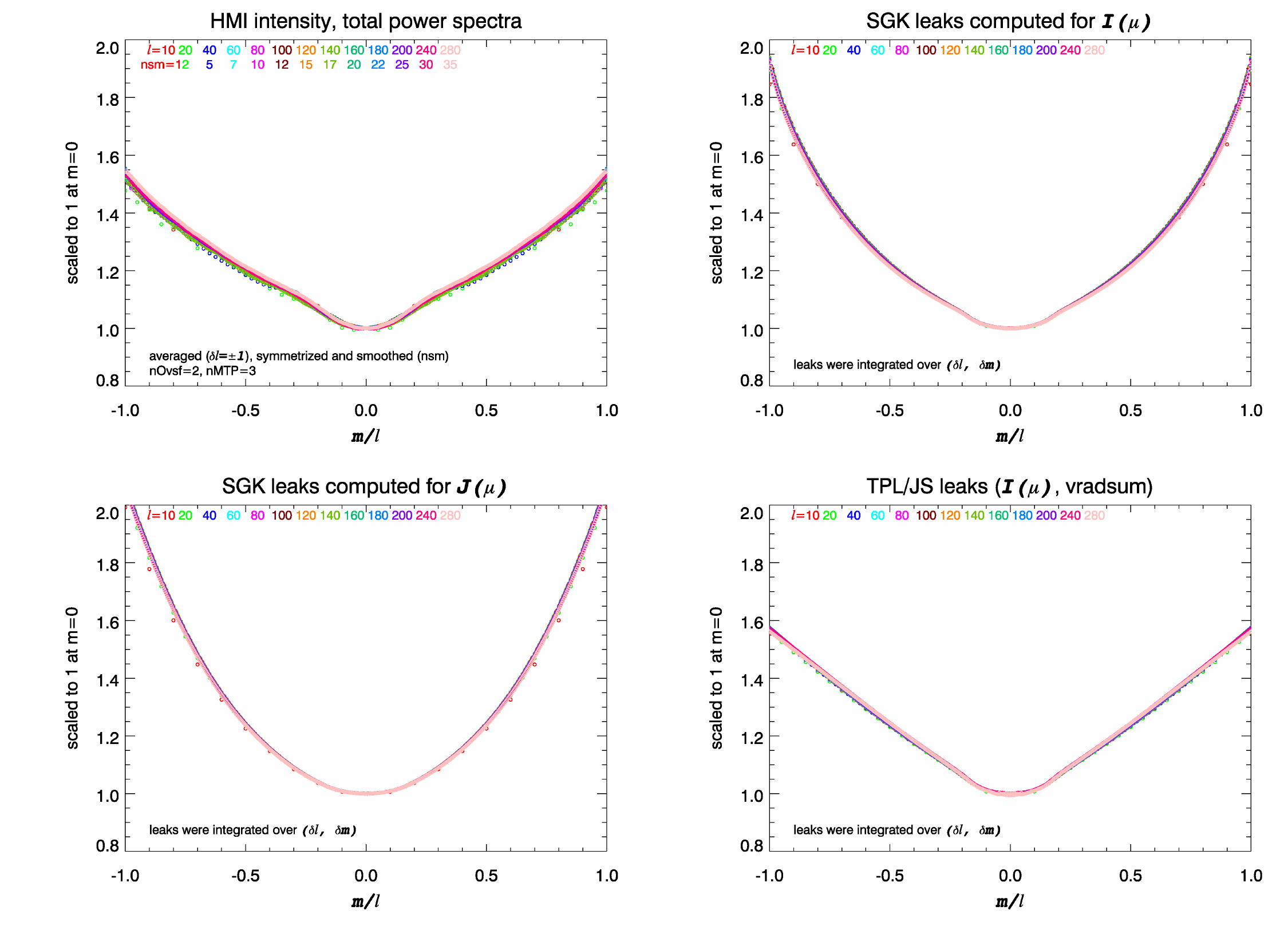}
\end{center}
\caption{Upper left panel: total power in the power spectra of the
  HMI intensity oscillation signal, for a set of degrees, $\ell= 10, 20, (20),
  200, (40), 280$, plotted as a function of the ratio $m/\ell$, and normalized
  to unity at $m=0$. The total power was smoothed in $m$ (as indicated by the
  $nsm$ key) and symmetrized with respect
  to $m/\ell$, and estimated using a $\delta\ell=\pm1$ range in $\ell$ to
  increase the significance of the derived profiles.
  The other three panels show the sum of the leaks, for the same set of
  degrees, also plotted as a function of the ratio $m/\ell$, and normalized to
  unity at $m=0$, \ie,$\bar{Q}_{\ell,m}^{\rm Tot} = \frac{1}{Q_{N}}
  \Sigma_{\delta\ell,\delta m}\,C^2_{\ell,m}(\delta\ell,\delta m)$. Each of
  these three panels corresponds to leakage estimates based on different
  $J(\mu)$ or $I(\mu)$ profiles.
\label{fig:valid-1a}}
\end{figure}

\Clearpage
\begin{figure}
\begin{center}
\includegraphics[width=.975\textwidth]{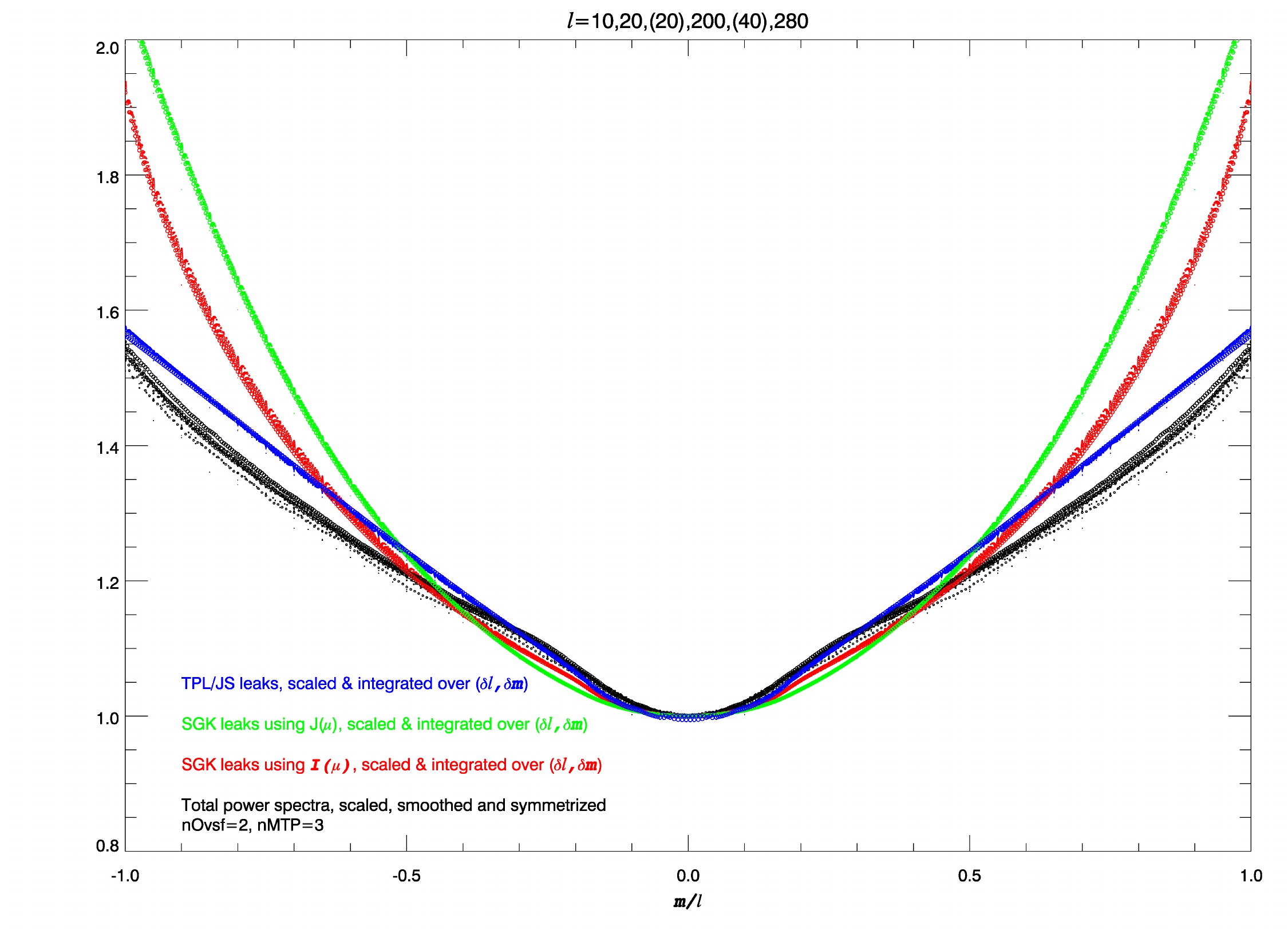}
\end{center}
\caption{Direct comparison of the profiles shown in Fig.~\ref{fig:valid-1a},
  with the size of the symbol proportional to $\ell$. Note that none of the
  leakage computations match the observed total power, nor do they duplicate
  correctly the distinctive ``{kink}'' near $m/\ell=0.25$ seen in the power
  profiles, although the case computed by me, using $I(\mu)$, displays a hint
  of a qualitatively similar {kink}.
\label{fig:valid-1b}}
\end{figure}

\Clearpage
\begin{figure}
\begin{center}
\includegraphics[width=.975\textwidth]{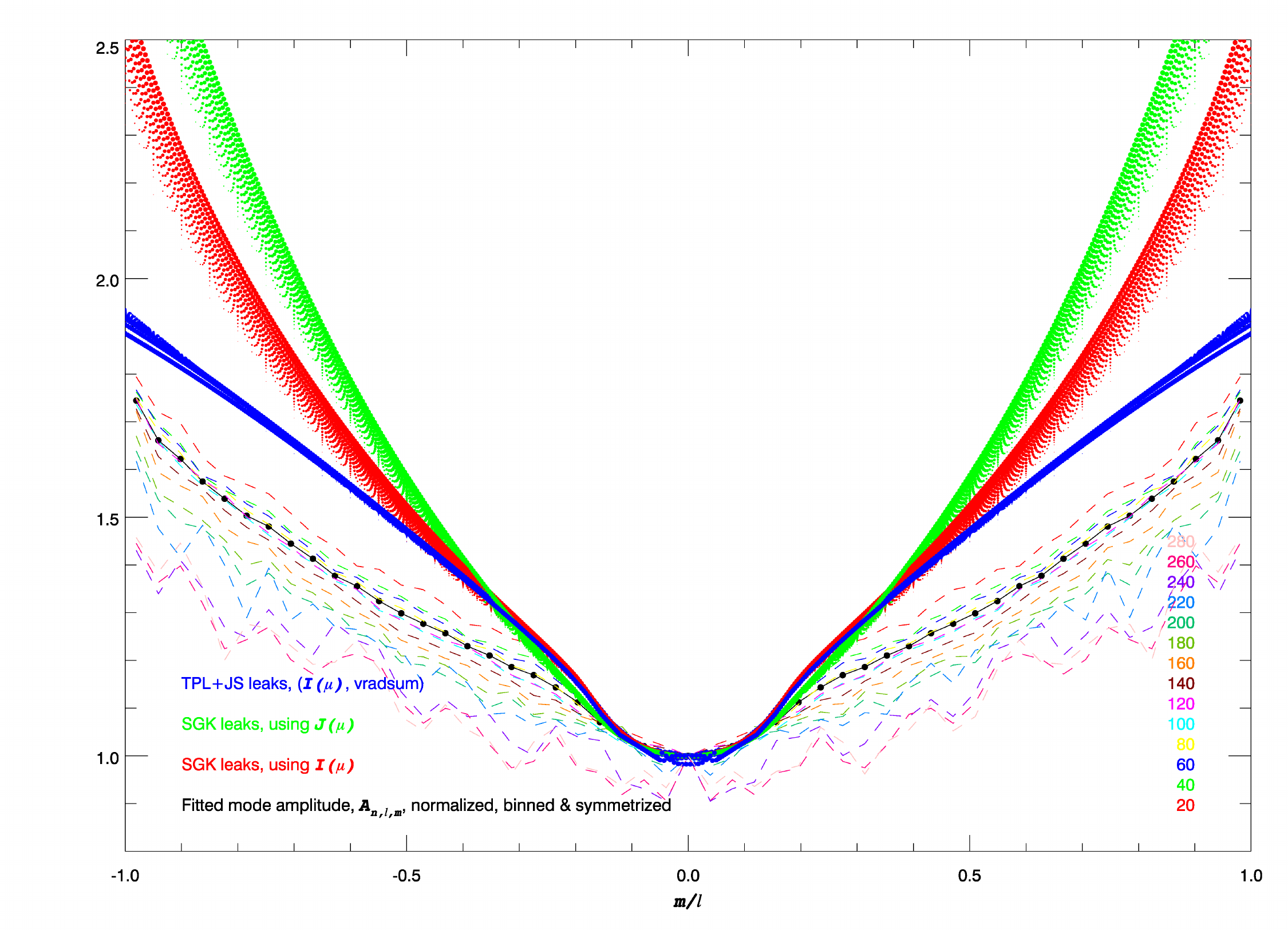}
\end{center}
\caption{Comparison, like the one shown in Fig.~\ref{fig:valid-1b}, but using
  the fitted mode power amplitude, $A_{n,\ell,m}$ to estimate the normalized
  and symmetrized observed power distribution profile with respect to $m/\ell$
  and shown as the connected black dots. The normalized values
  $\bar{Q}_{\ell,m}=\frac{C^2_{\ell,m}(0,0)}{C^2_{\ell,0}(0,0)}$,\ie, no
  summation on $(\delta\ell,\delta m)$, are shown with colored dots, with
  their size being proportional to $\ell$. The dash colored lines correspond
  to estimates of the observed power distribution profile, derived from
  measured $A_{n,\ell,m}$ but restricted to a given range in $\ell$ centered
  around a target $\ell$.
  Note that again none of the leakage computations match the mode profile
  power amplitude, nor do they duplicate correctly the distinctive ``{kink}''
  seen in the observations, although the case computed by me and by the
  Stanford group, both using $I(\mu)$, displays a hint of a qualitatively
  similar {kink}.
\label{fig:valid-2}}
\end{figure}

\Clearpage
\begin{figure}
\begin{center}
\includegraphics[width=.975\textwidth]{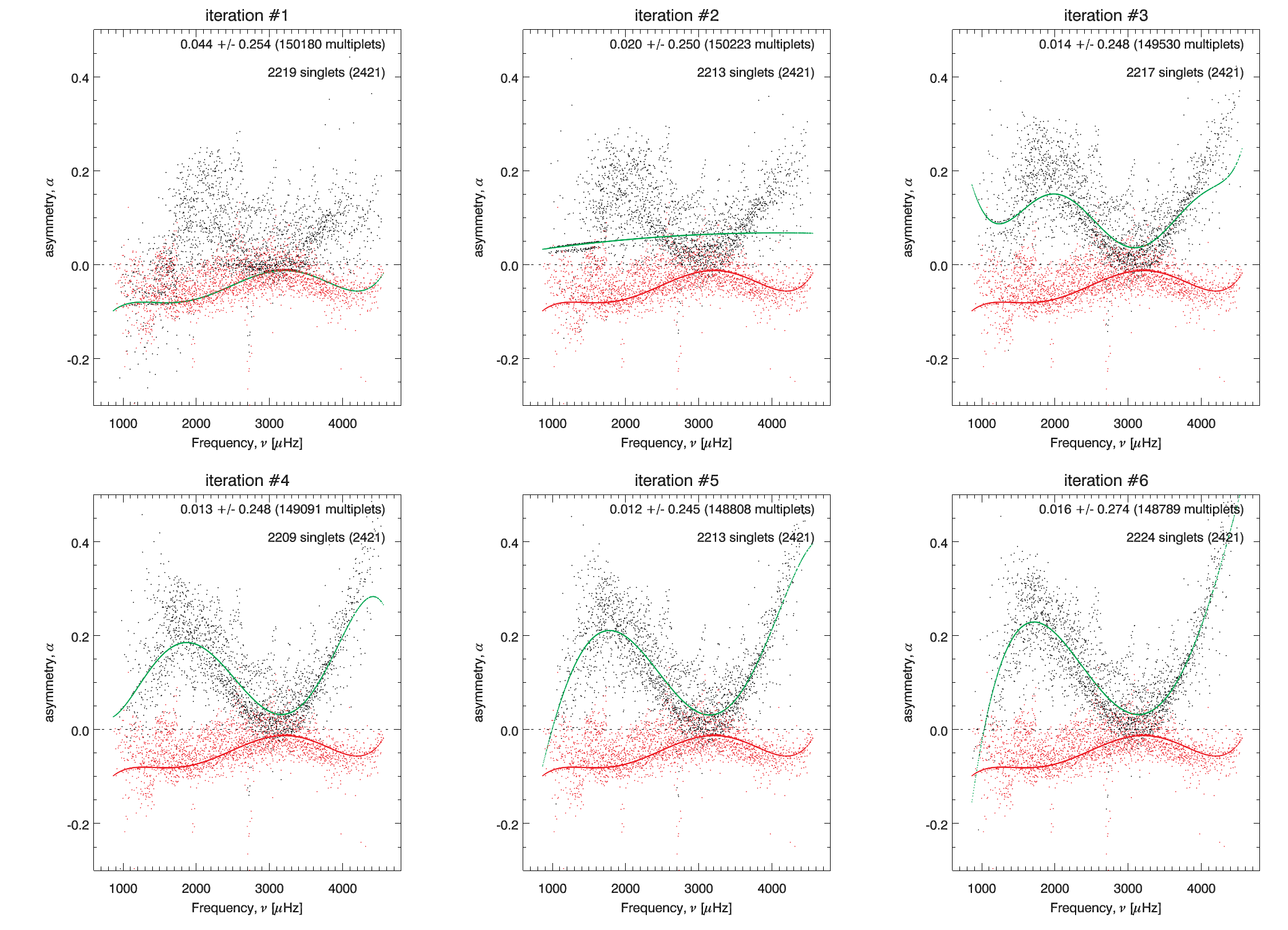}
\end{center}
\caption{Values of the seed. \ie, initial guess, and fitted asymmetry,
  plotted as a function of frequency, for each iteration used in the
  refinement of the seed asymmetry values. The red dots are values of
  $\alpha^V_{n,\ell}$ resulting from fitting velocity observations, the black
  dots are values of $\alpha^I_{n,\ell}$ resulting from fitting intensity
  observations at each successive iteration. The red curves show the seed
  asymmetry $\alpha_{n,\ell}^{sV}$ used for velocity, the green curves show
  the seed asymmetry $\alpha_{n,\ell}^{sI}$ for intensity at each
  iteration. The mean and standard deviation of the changes in the fitted
  frequency values at each iteration are indicated in each panel.
  Note how even with initial negative values for the asymmetry, the resulting
  fitted asymmetries become mostly positive at the first iteration (upper left
  panel).
\label{fig:seed-alpha}}
\end{figure}

\Clearpage
\begin{figure}
\begin{center}
\includegraphics[width=.925\textwidth]{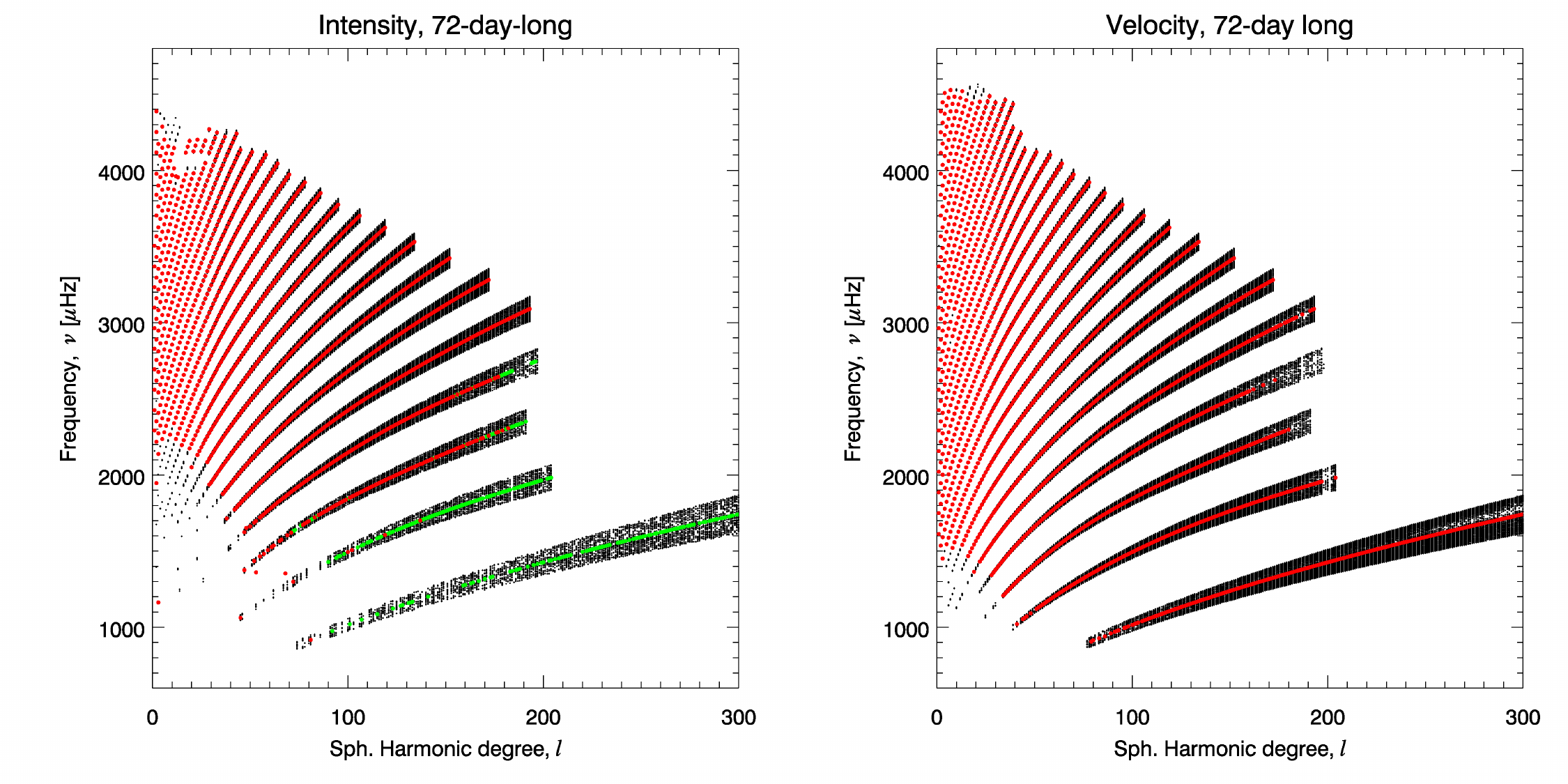}
\includegraphics[width=.925\textwidth]{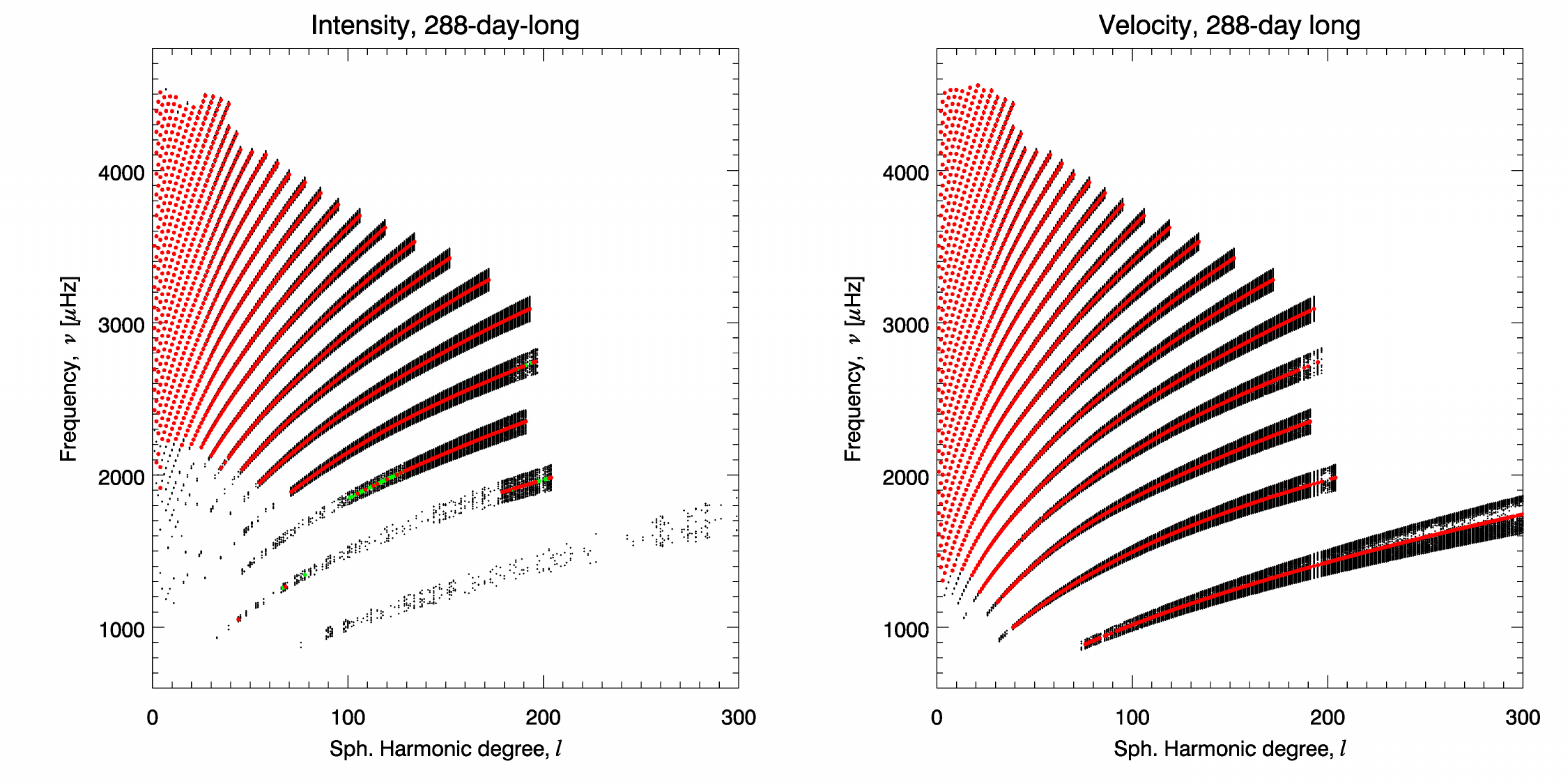}
\end{center}
\caption{Coverage in the $\ell-\nu$ plane of the fitted modes. Black dots show
  singlets, red dots multiplets and green dot multiplets resulting from a less
  restrictive rule for the conversion of singlets to multiplets (see
  explanation in the text). The top two panels correspond to fitting one
  72-day long time series, the bottom two panels to fitting one 288-day long
  time series. Panels on the left correspond to intensity observations, panels
  on the right to co-eval velocity observations.
  Note the reduced success rate in fitting intensity observations, especially
  for the low-order, low-frequency modes. Note also, the counter-intuitive
  higher success rate for fitting f-mode singlets for the 72-day long time
  series than for the 288-day long one, when using intensity observations.
\label{fig:lnu}}
\end{figure}

\Clearpage
\begin{figure}
\begin{center}
\includegraphics[width=.975\textwidth]{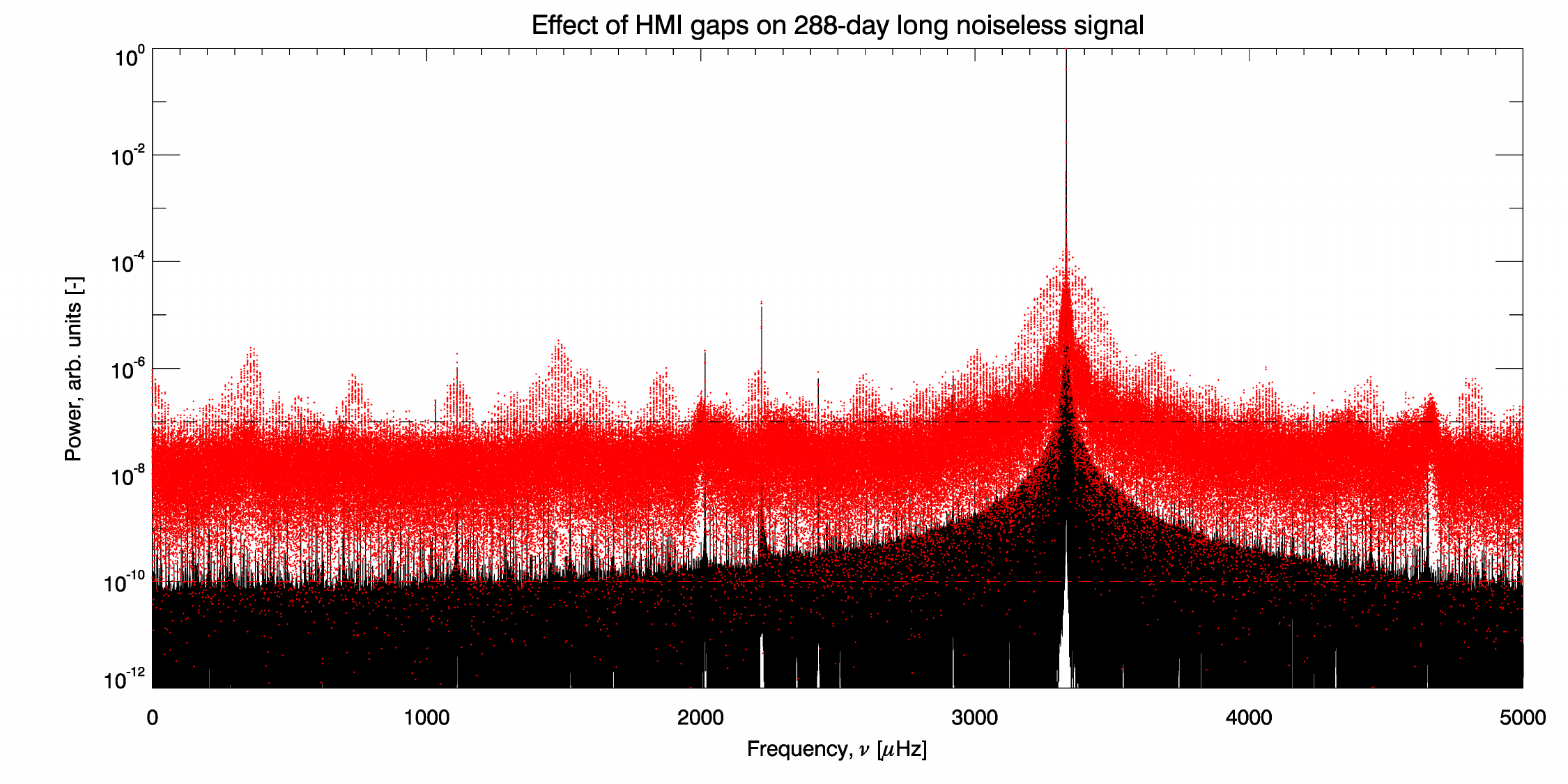}
\end{center}
\caption{Effect of gaps on a noiseless data set. The black curve is the power
  spectrum of a simple sine wave, sampled every 45 seconds for 288 days. The
  red curve is that same sine wave but with values set to zero at times when
  HMI observations are missing for the 288-day long time series analyzed. The
  introduction of gaps scatters power and raises the background levels
  considerably, but uniformly with respect to frequency, compared to the
  gap-less noiseless case.
\label{fig:gap-noise}}
\end{figure}

\Clearpage
\begin{figure}
\begin{center}
\includegraphics[width=.49\textwidth]{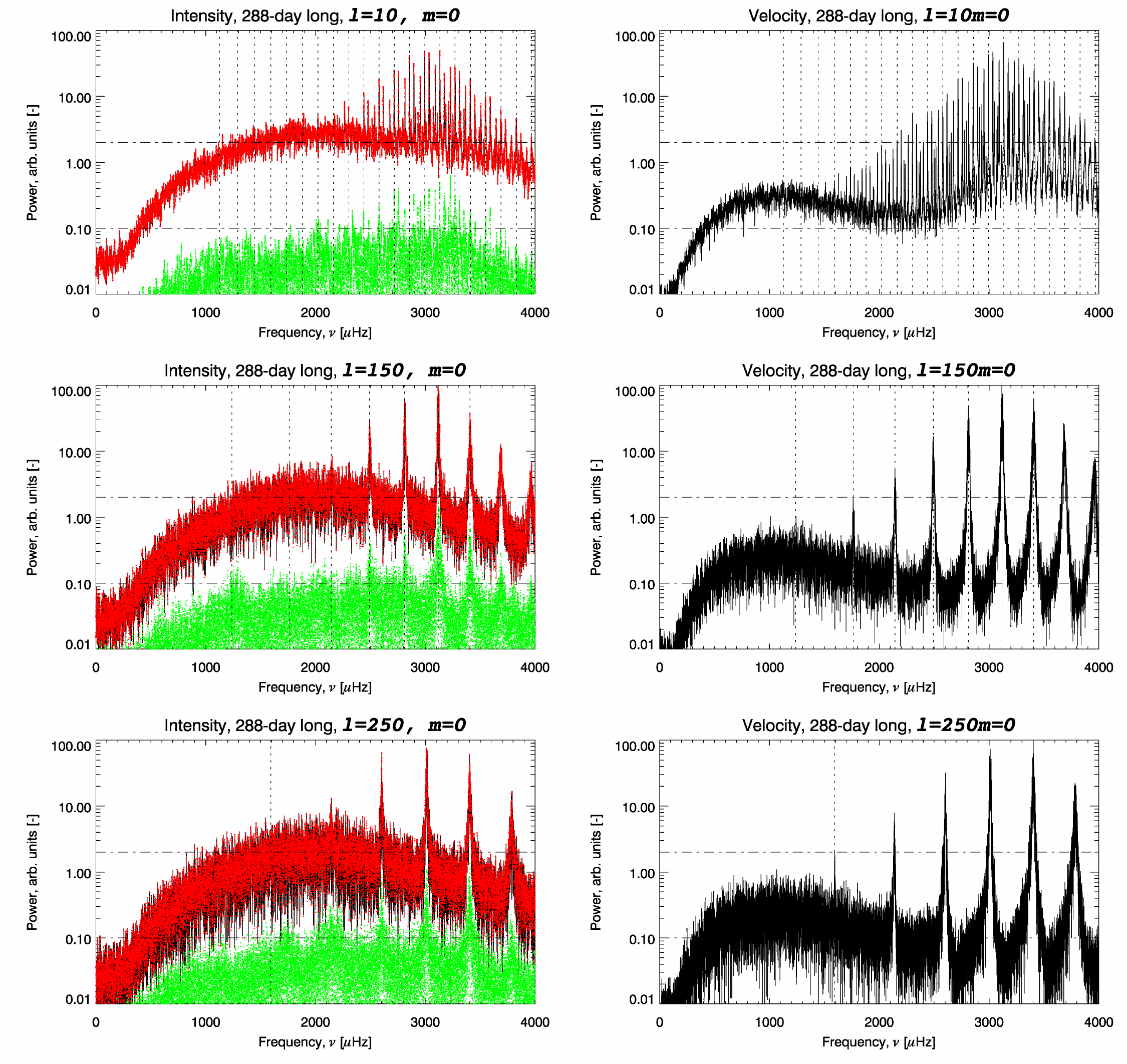}
\includegraphics[width=.49\textwidth]{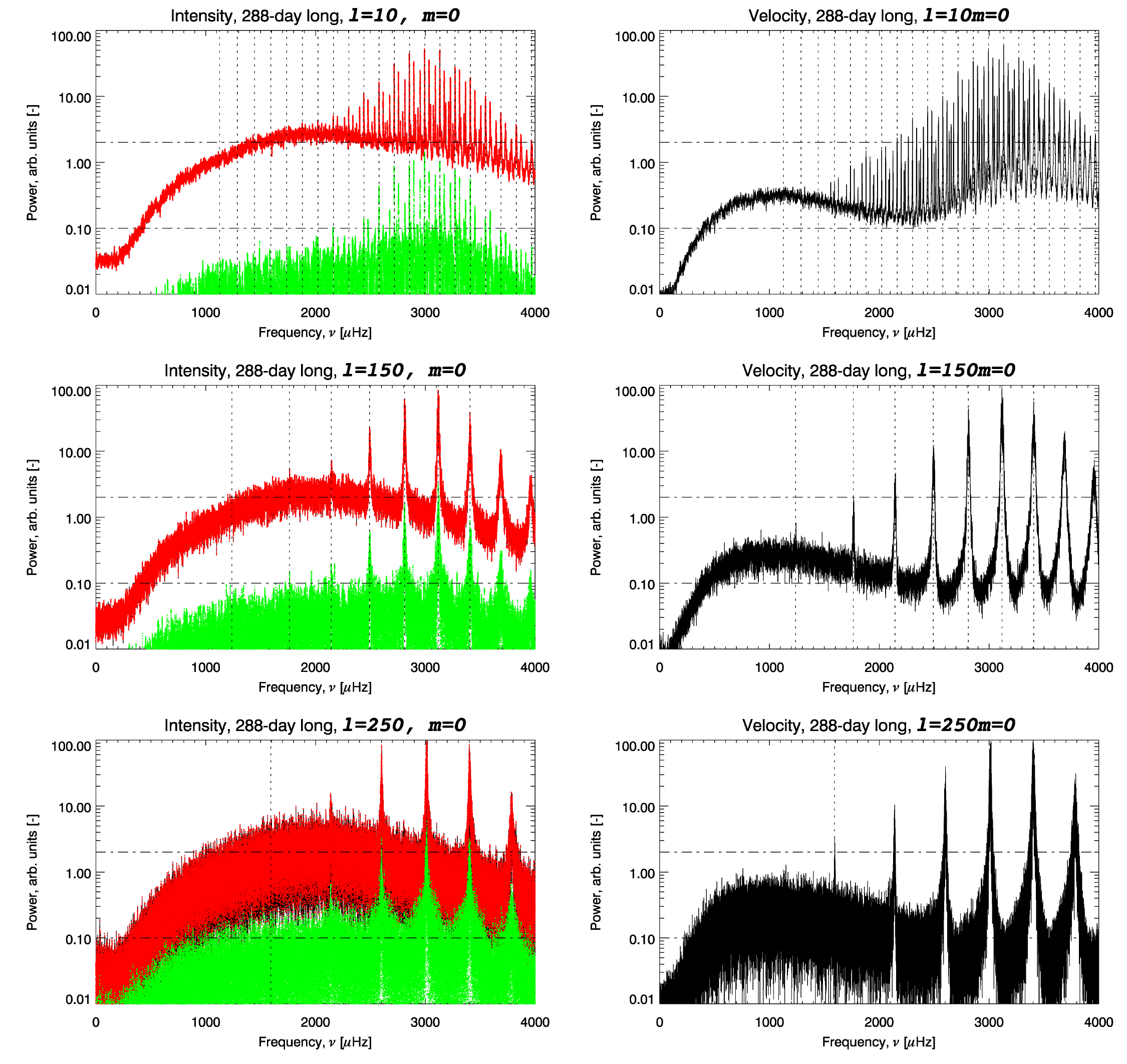}
\end{center}
\caption{Examples of sine multi-taper zonal ($m=0$) power spectra computed
  using 72-day long (six leftmost panels) or 288-day long (six rightmost
  panels) time series.  The panels in the first and third columns show power
  spectra of intensity observations derived from raw or gap-filled time series (red
  and black curves respectfully). The green curves are the difference between
  the spectra computed using raw or gap-filled time series. The second and
  forth columns show power spectra of gap-filled co-eval velocity
  observations. The vertical lines indicate the location of the modes being
  fitted. Each row corresponds to a different value of $\ell$
  ($\ell=10,150,250$, top to bottom respectfully), the horizontal lines are
  drawn as fiducial lines to mark the background level around 2 mHz.
%
%
%
\label{fig:gap-filled}}
\end{figure}

\Clearpage
\begin{figure}
\begin{center}
\includegraphics[width=.49\textwidth]{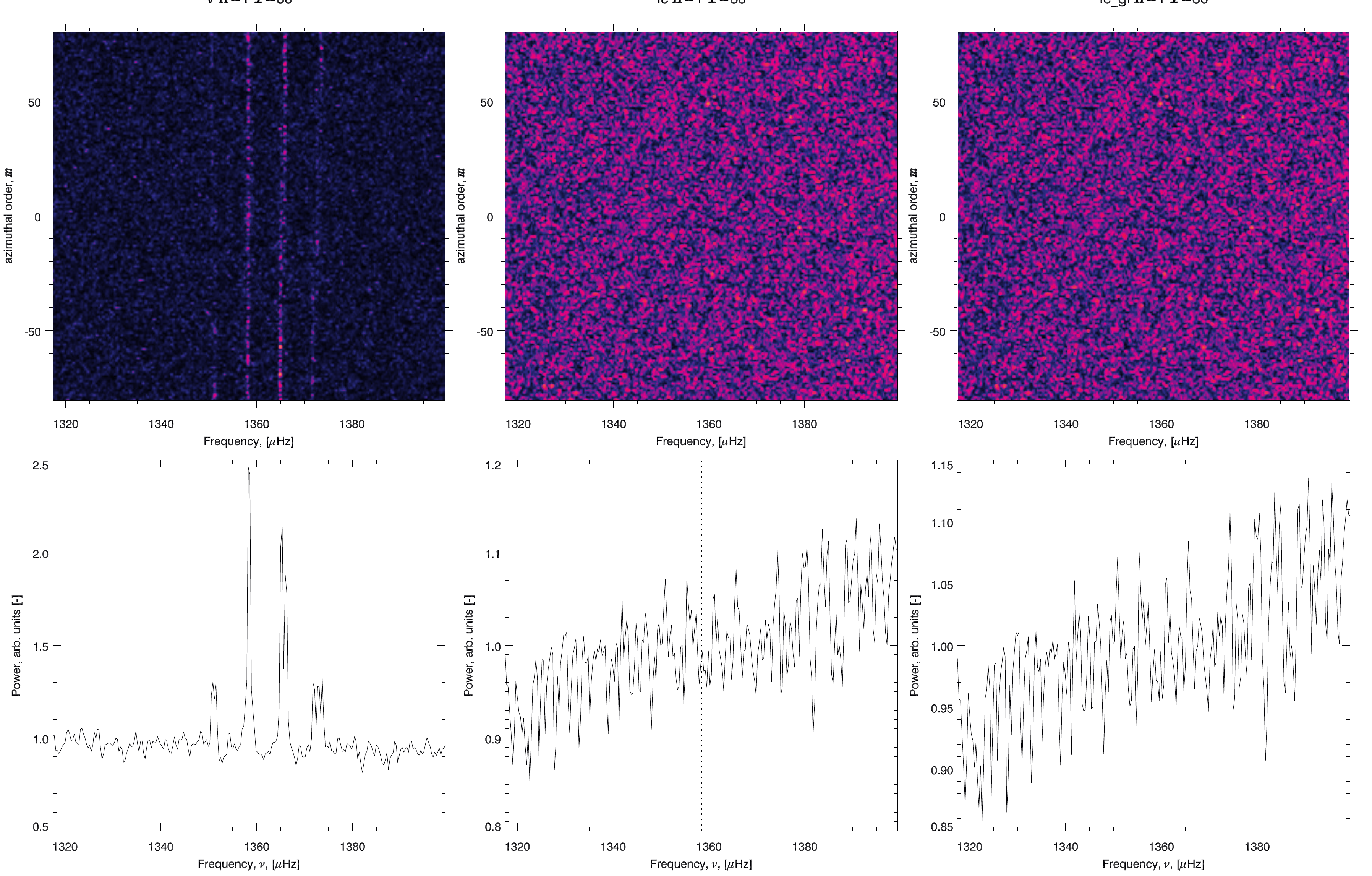}
\includegraphics[width=.49\textwidth]{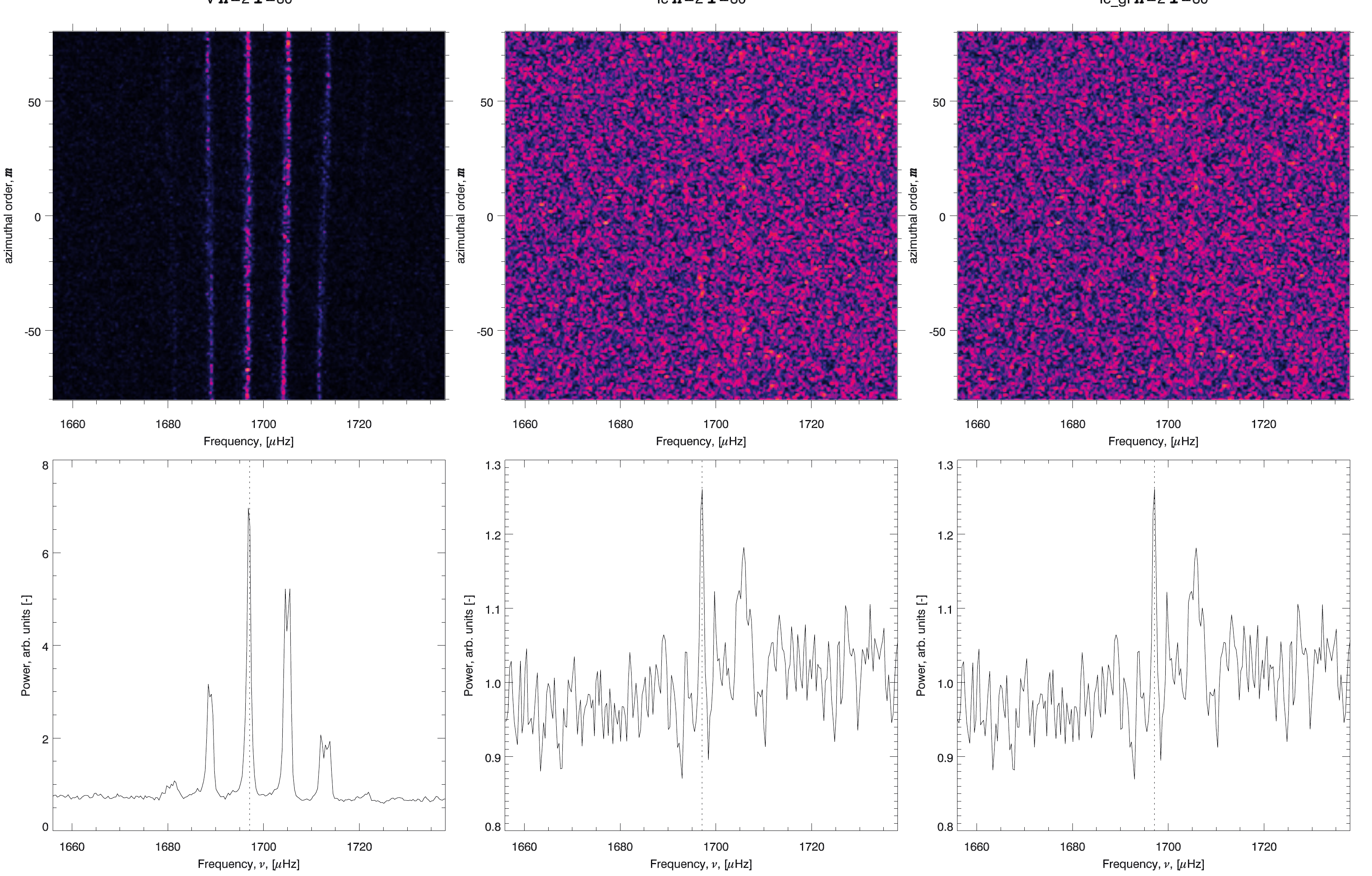}
\includegraphics[width=.49\textwidth]{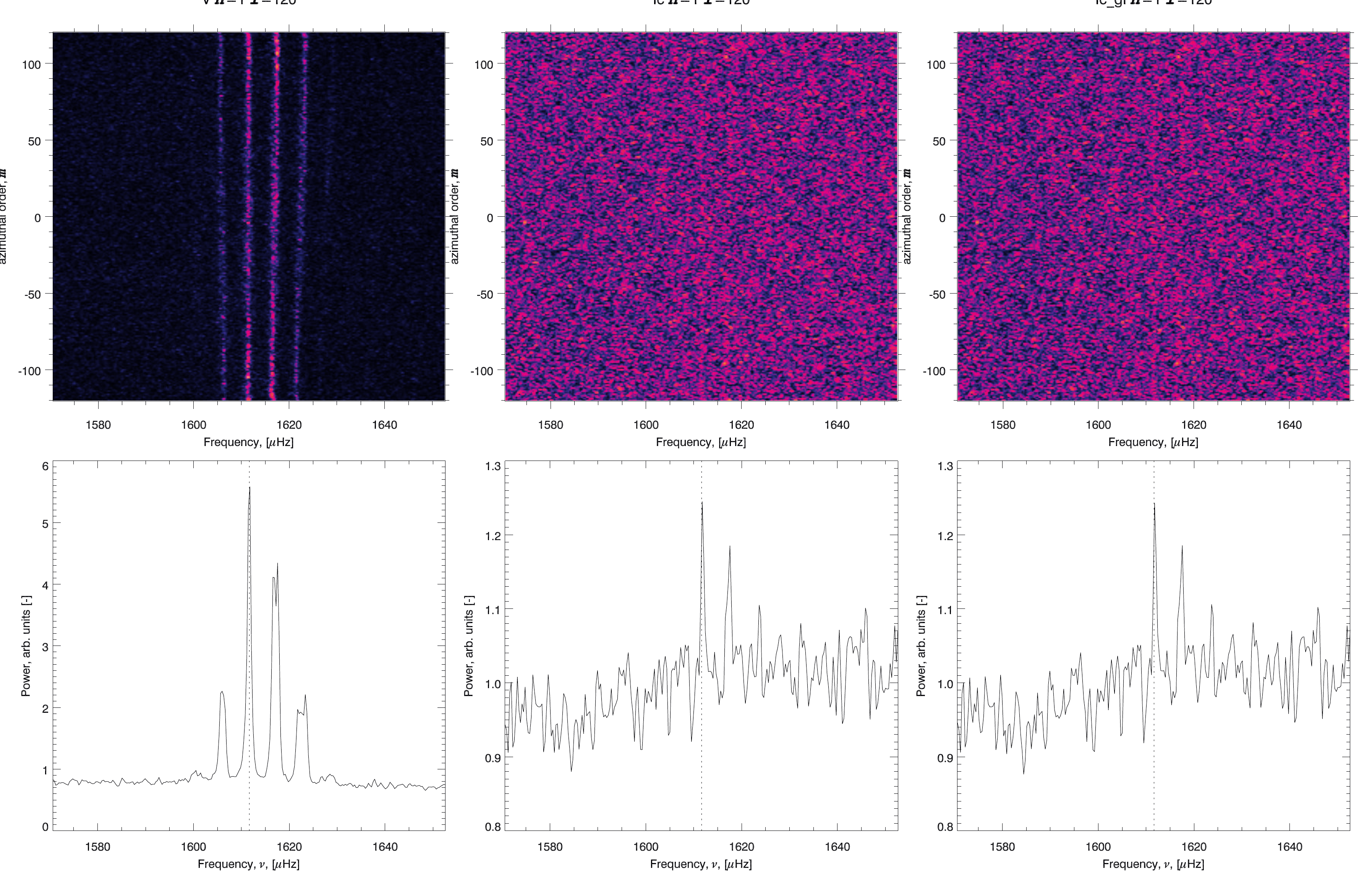}
\includegraphics[width=.49\textwidth]{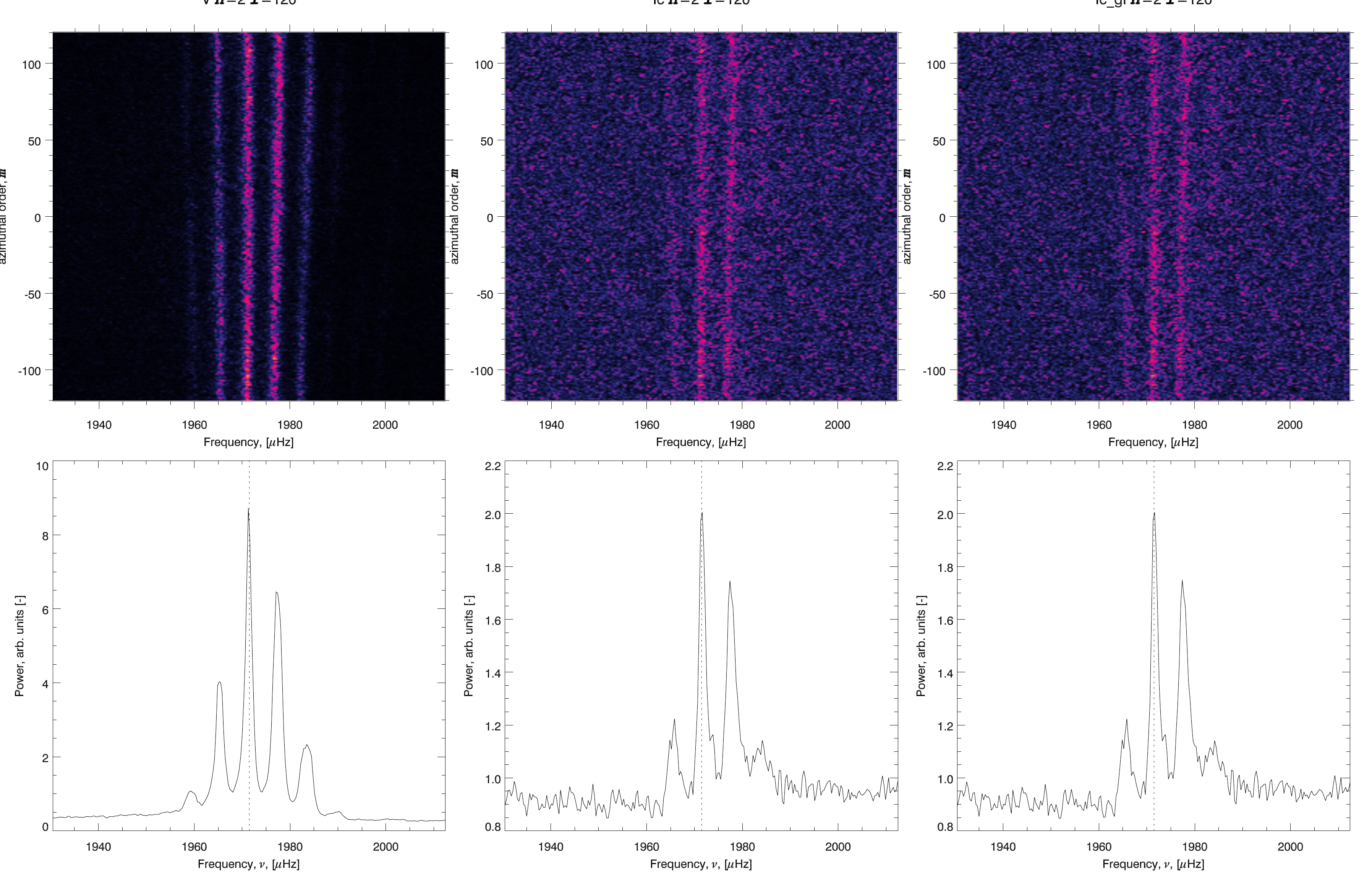}
\end{center}
\caption{Stacked sections of power spectra, shown as a function of $\nu$ and
  $m$, and the corresponding $m$-averaged spectra, centered around a set of
  four modes for $(n,\ell) = (1,80), (2,80), (1,120)$ and $(2,120)$, computed
  using a 72-day long time series. The vertical lines indicate the mode
  frequency. Each set of six panels shows in the top row the stacked spectra,
  in the bottom row the corresponding $m$-averaged spectrum, and from left to
  right, spectra computed from the gap-filled velocity, the raw ({\em i.e.},
  with gaps) intensity and the gap-filled intensity co-eval time series.
  Stacked sections of power spectra are sections of spectra centered on the
  mode singlet frequency, $\nu_{n,\ell,m}$, computed using a very good
  estimate of the mode frequency and frequency splitting to offset in
  frequency the spectrum for each $m$ so as to co-align the target modes.
\label{fig:show-spc-72d}}
\end{figure}

\Clearpage
\begin{figure}
\begin{center}
\includegraphics[width=.49\textwidth]{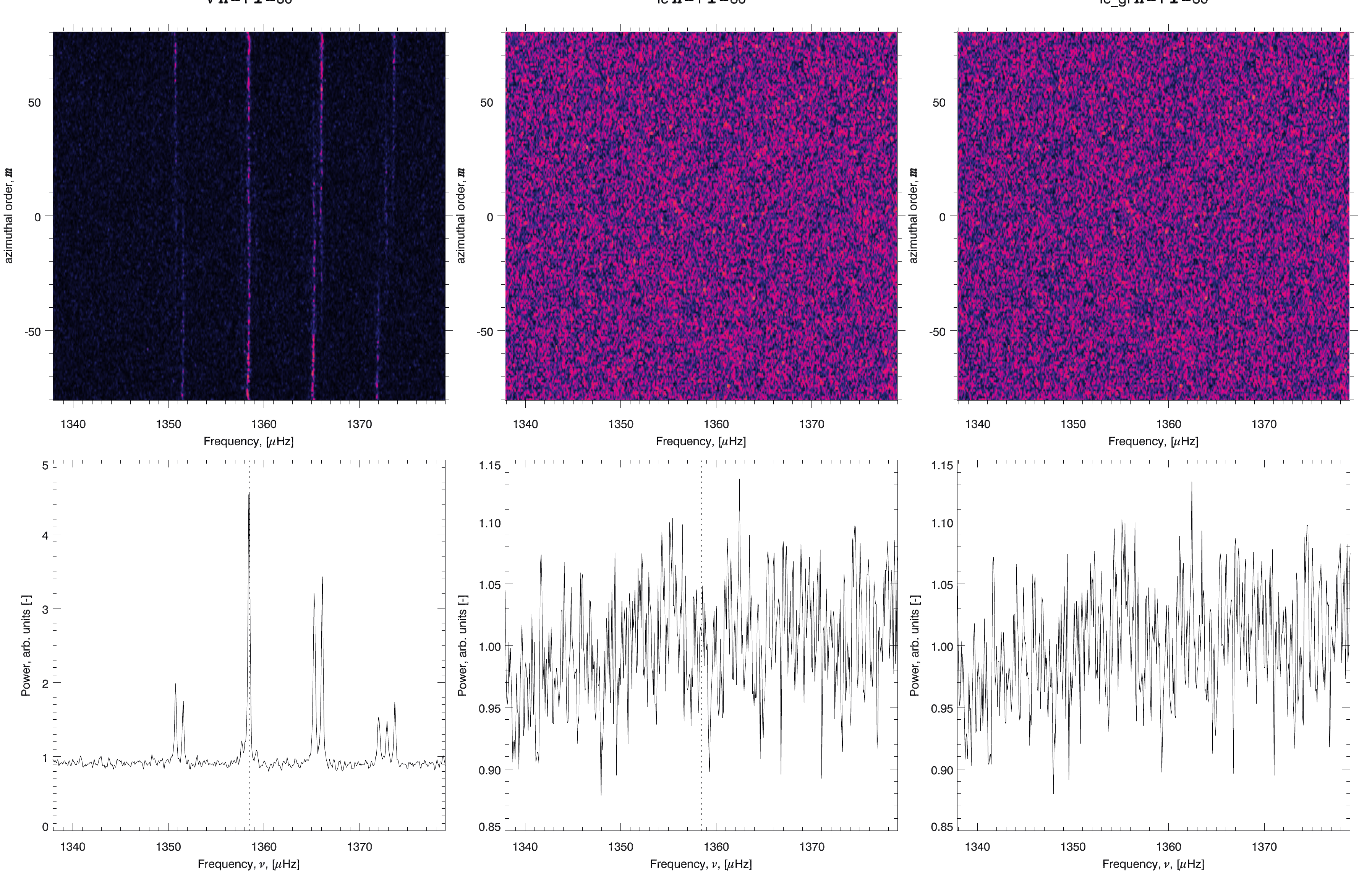}
\includegraphics[width=.49\textwidth]{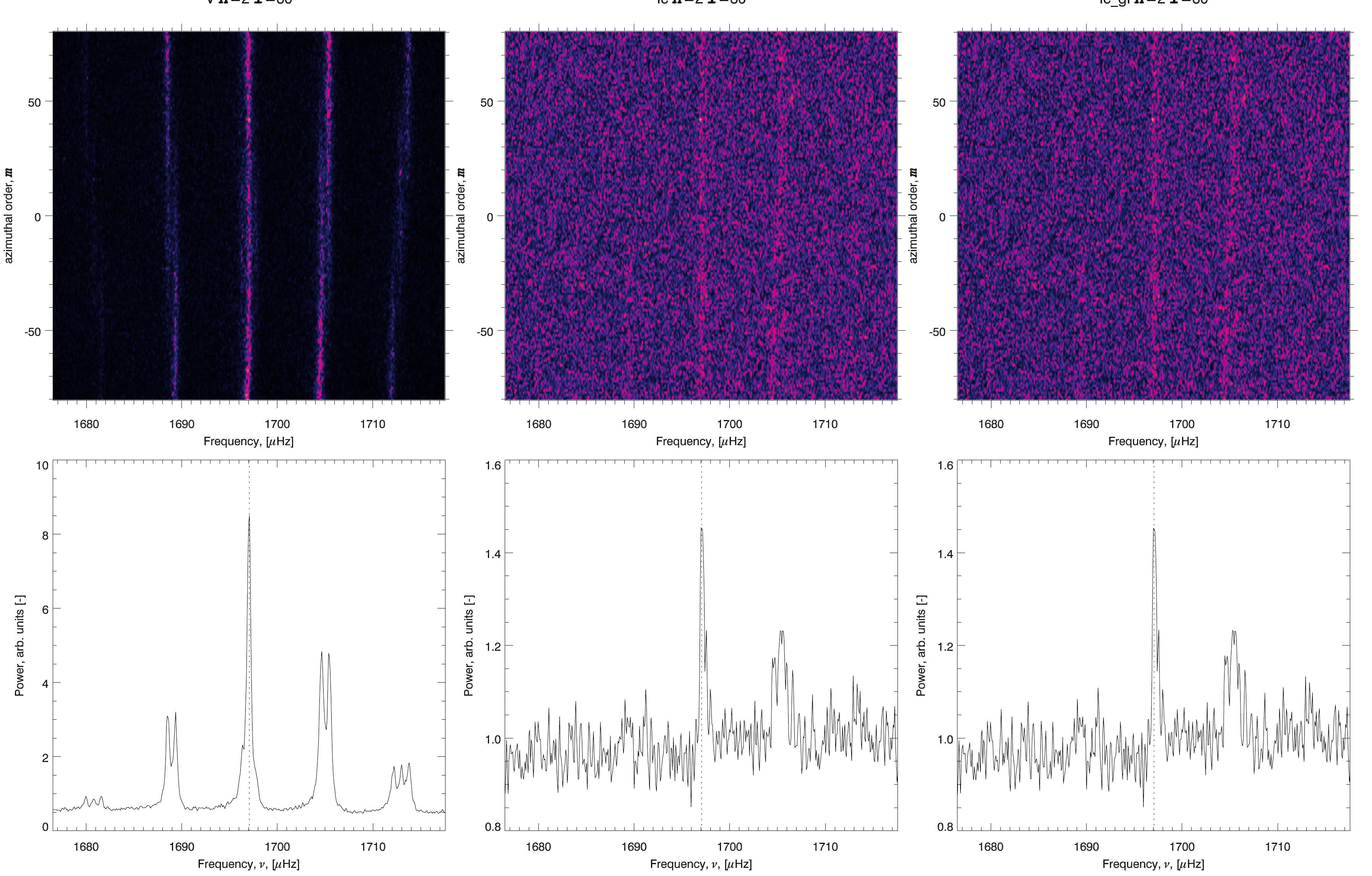}
\includegraphics[width=.49\textwidth]{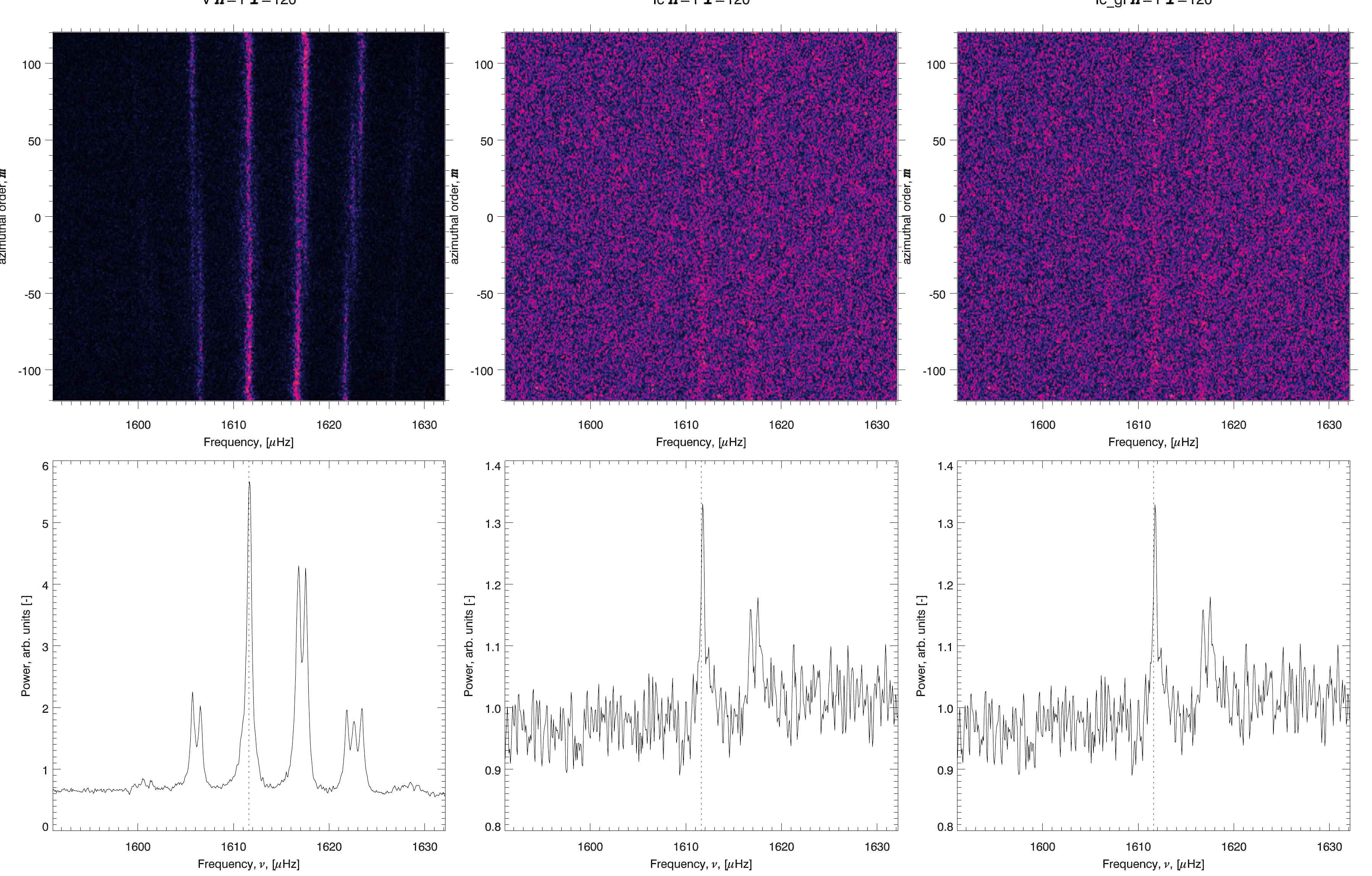}
\includegraphics[width=.49\textwidth]{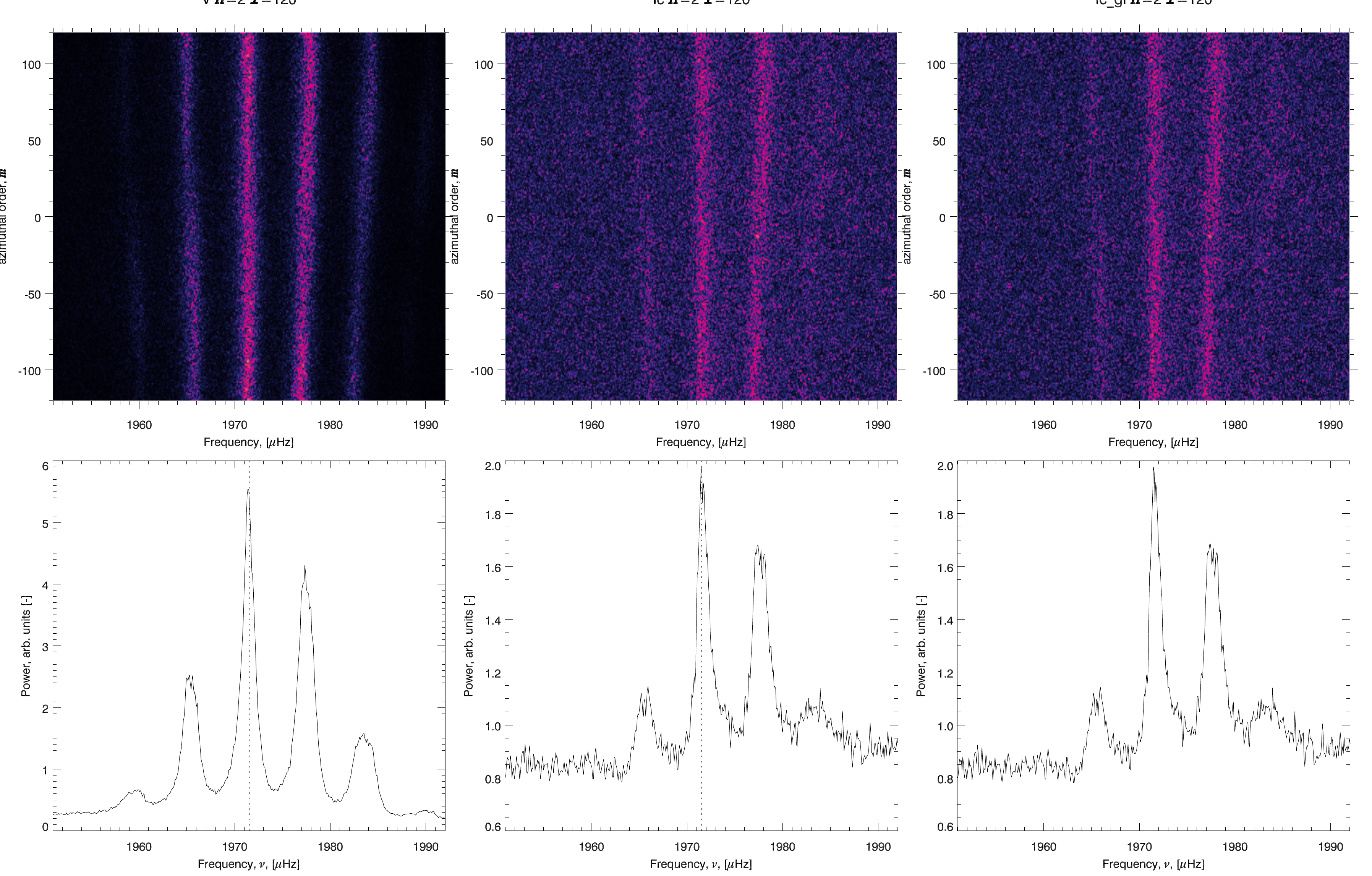}
\end{center}
\caption{Stacked and $m$-averaged spectra, like in
  Fig.~\ref{fig:show-spc-72d}, but when using a 288-day long time series. The
  $(n,\ell) = (2,80)$ and $(1,120)$ modes become barely visible in intensity
  when using a longer time series.
\label{fig:show-spc-288d}}
\end{figure}

\Clearpage
\begin{figure}
\begin{center}
\includegraphics[width=.77\textwidth]{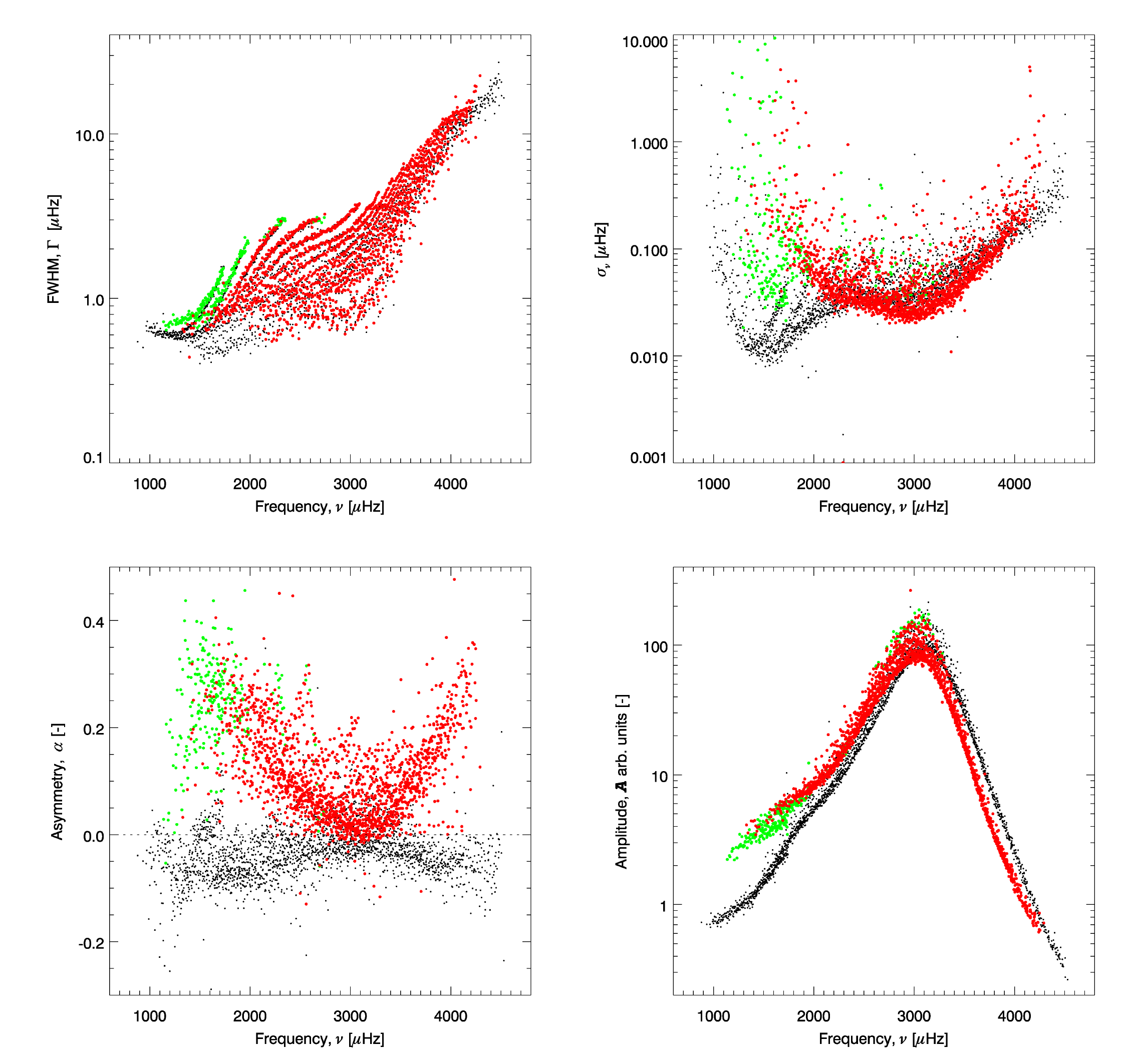}
\end{center}
\caption{Mode characteristics derived from fitting a 72-day long time series,
  after converting singlets to multiplets. The panels show, as a function of
  frequency, the modes FWHM, $\Gamma$, frequency uncertainty, $\sigma_{\nu}$,
  asymmetry, $\alpha$, and, the mean mode amplitude,
  $\bar{A}_{n,\ell}=\frac{1}{N_m}\Sigma_{m} A_{n,\ell,m}$.  The red and green
  circles correspond to fitting intensity observations, with the green circles
  resulting from a less restrictive constraint in the conversion of singlets
  to multiplets (see text and Fig.~\ref{fig:lnu}), the black dots corresponding
  to results from fitting co-eval velocity time series.
  Except for the low-order low-frequency modes, the FWHM and the frequency
  uncertainties derived using either velocity or intensity agree quite
  well. The asymmetry when fitting intensity observations is both of opposite
  sign to the asymmetry for velocity but also larger in magnitude. The mode
  power amplitude distribution, while peaking at the same frequency and being
  overall similar, shows a distinctive different distribution with frequency
  when fitting intensity rather than velocity observations.
  The green circles, resulting from estimating the multiplets using relaxed
  rules, appear to be consistent with their corresponding values derived from
  velocity but show a larger uncertainty. This in itself is not surprising
  since they are derived from fewer individually fitted singlets.
\label{fig:results-72d}}
\end{figure}

\Clearpage
\begin{figure}
\begin{center}
\includegraphics[width=.77\textwidth]{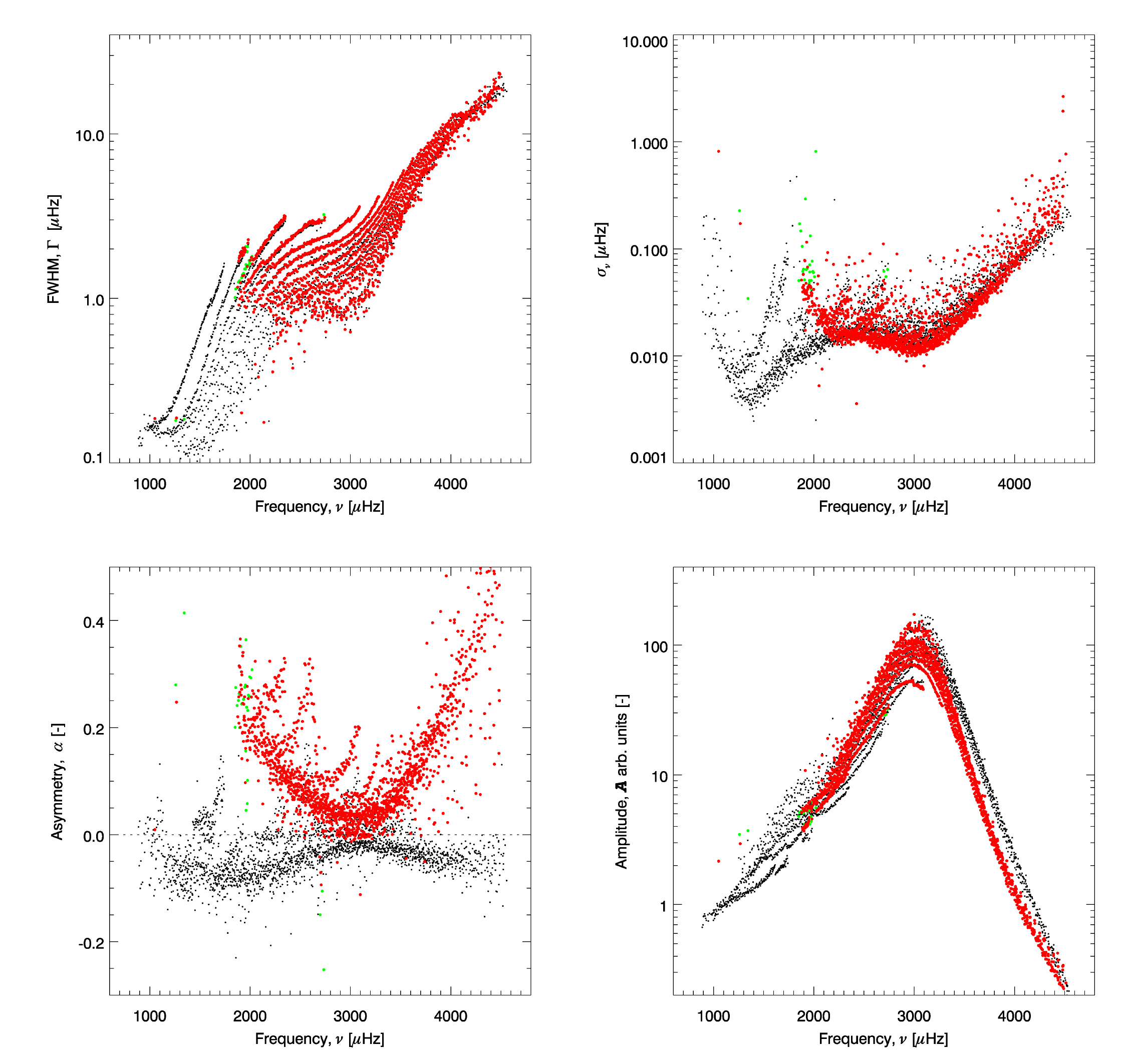}
\end{center}
\caption{Mode characteristics derived from fitting a 288-day long time series,
  after converting singlets to multiplets, using the same representation as
  the one used in Fig.~\ref{fig:results-72d}.
  Again, like for the 72-day long results, the frequency uncertainties derived
  using either velocity or intensity observations agree quite well. The
  asymmetry for intensity observations is both of opposite sign to the
  asymmetry for velocity observation and larger in magnitude. The mode power
  amplitude distributions derived from intensity or velocity observations
  also peak at the same frequency but are also more similar than for the
  72-day long case.  The frequency uncertainties are, as expected, reduced by
  a factor $\sqrt{{288}/{72}}=2$ when compared to values obtained using a
  shorter time series.
  By contrast to the 72-day long results, very few individual singlets were
  fitted for modes with $\nu < 1800$ $\mu$Hz and $\Gamma < 0.8$ $\mu$Hz.
\label{fig:results-288d}}
\end{figure}

\Clearpage
\begin{figure}
\begin{center}
\includegraphics[width=.85\textwidth]{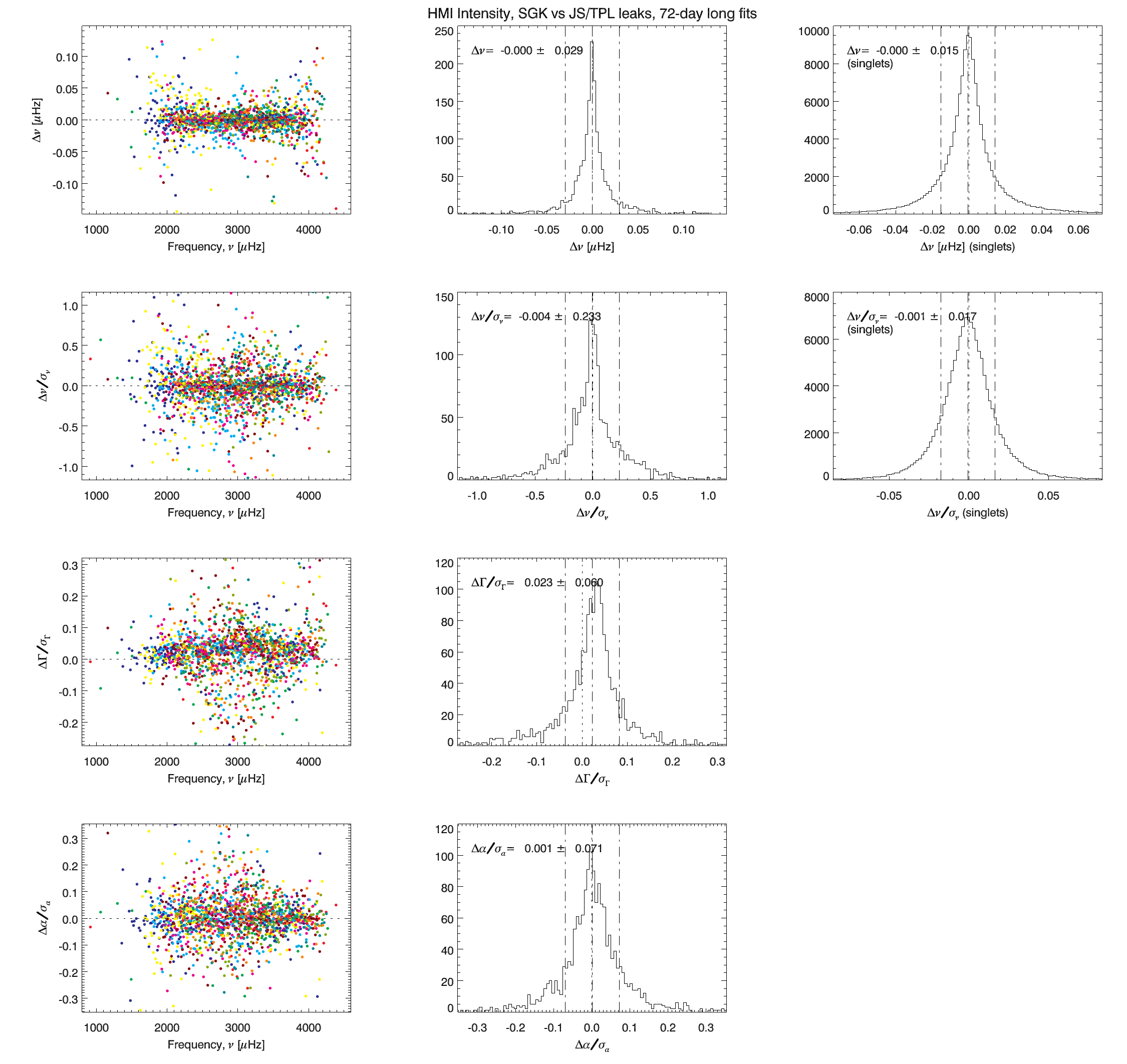}
\end{center}
\caption{Comparison of mode characteristics derived from fitting a 72-day long
  time series, after converting singlets to multiplets, and for singlets,
  using two different leakage matrix computations, namely mine based on
  $I(\mu)$ and the one computed by the Stanford group. The panels show, top to
  bottom, raw and scaled frequency differences for multiplets and singlets,
  and FWHM and asymmetry scaled differences.  The panels in the leftmost
  column show the multiplets' differences with colors corresponding to the
  modes' order, $n$. The panels in the middle column show the histogram
  distribution of the differences for the multiplets, while the panels in the
  rightmost column show the histogram distribution of the differences computed
  using singlets. Vertical lines are drawn at zero and at the mean plus or
  minus one standard deviation around the mean.
  Despite significant differences between the two leakage matrix coefficients
  (see, for example, Fig.\ref{fig:valid-2}), the resulting parameters show
  little differences, both in term of bias and spread.
%
%
  Only the FWHM differences show a non-negligible bias.
  \label{fig:compare-leakage-72d}}
\end{figure}

\Clearpage
\begin{figure}
\begin{center}
\includegraphics[width=.85\textwidth]{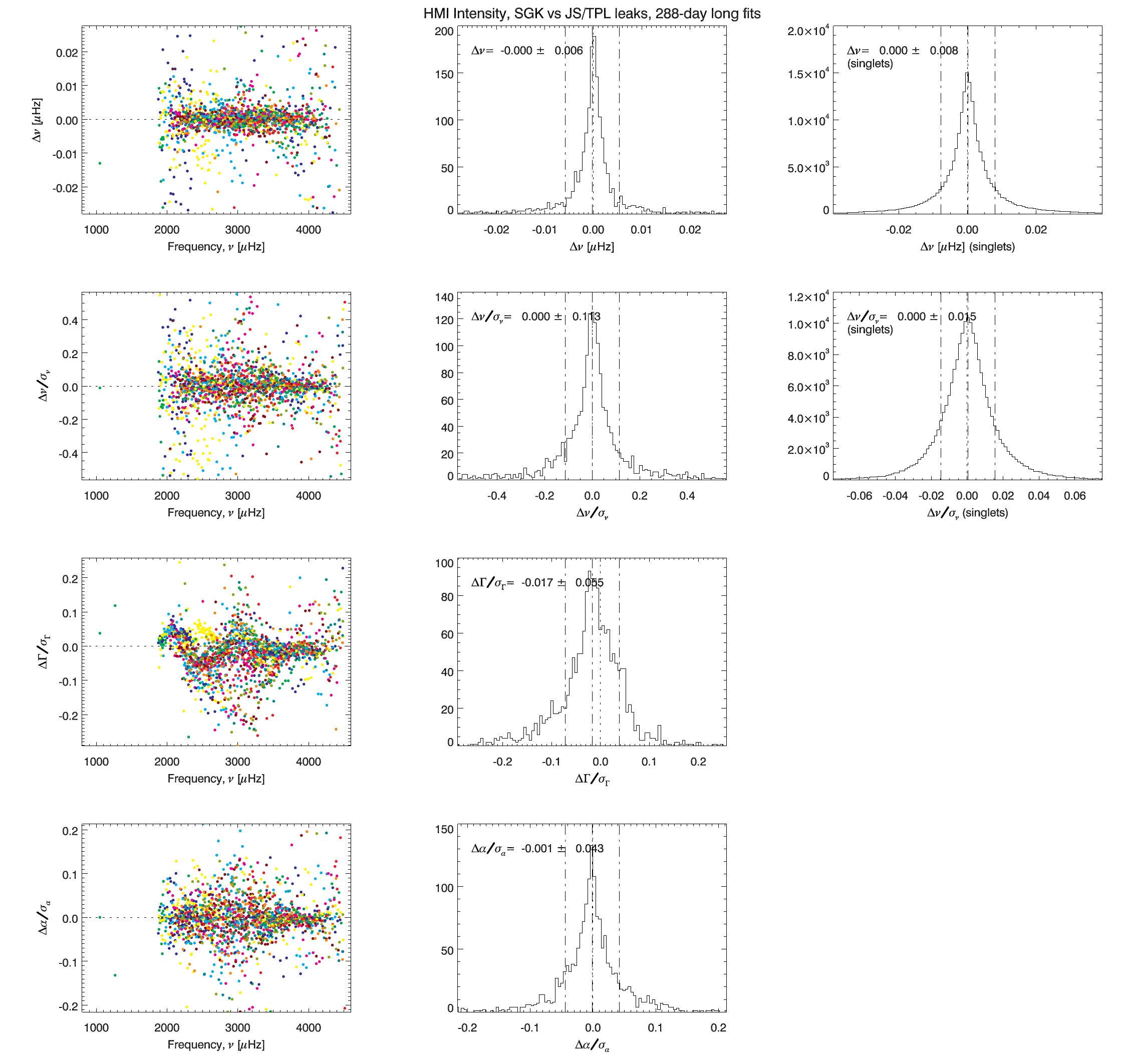}
\end{center}
\caption{Comparison of mode characteristics derived from fitting a 288-day
    long time series, as shown in Fig.~\ref{fig:compare-leakage-72d} for the
    72-day long case, and using the same two different leakage matrix
    computations. The resulting bias and spread remain small.
%
    Again, the FWHM differences show a non-negligible bias, that while small,
    show a hint of systematic distribution with order and frequency.
\label{fig:compare-leakage-288d}}
\end{figure}

\Clearpage
\begin{figure}
\begin{center}
\includegraphics[width=.85\textwidth]{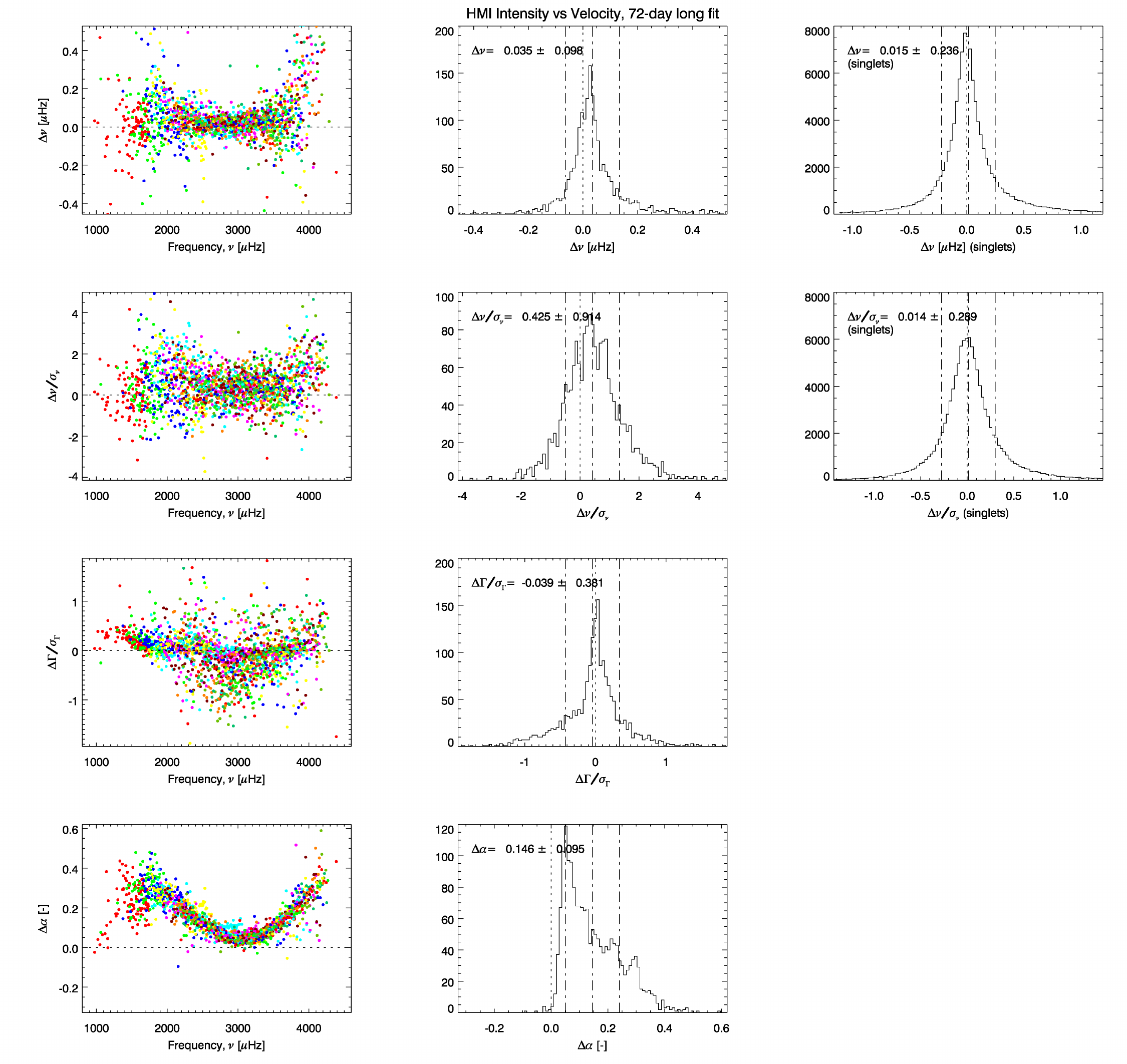}
\end{center}
\caption{Comparison of mode characteristics derived from fitting a 72-day long
  time series, as shown in Fig.~\ref{fig:compare-leakage-72d}, but resulting
  from fitting co-eval intensity and velocity time series. The means and
  standard deviation of the differences are negligible for frequencies,
  $\nu$, and FWHM, $\Gamma$.
  As expected the differences in asymmetries are large and show a clear and
  smooth dependence on frequency.
\label{fig:compare-vel-int-72d}}
\end{figure}

\Clearpage
\begin{figure}
\begin{center}
\includegraphics[width=.85\textwidth]{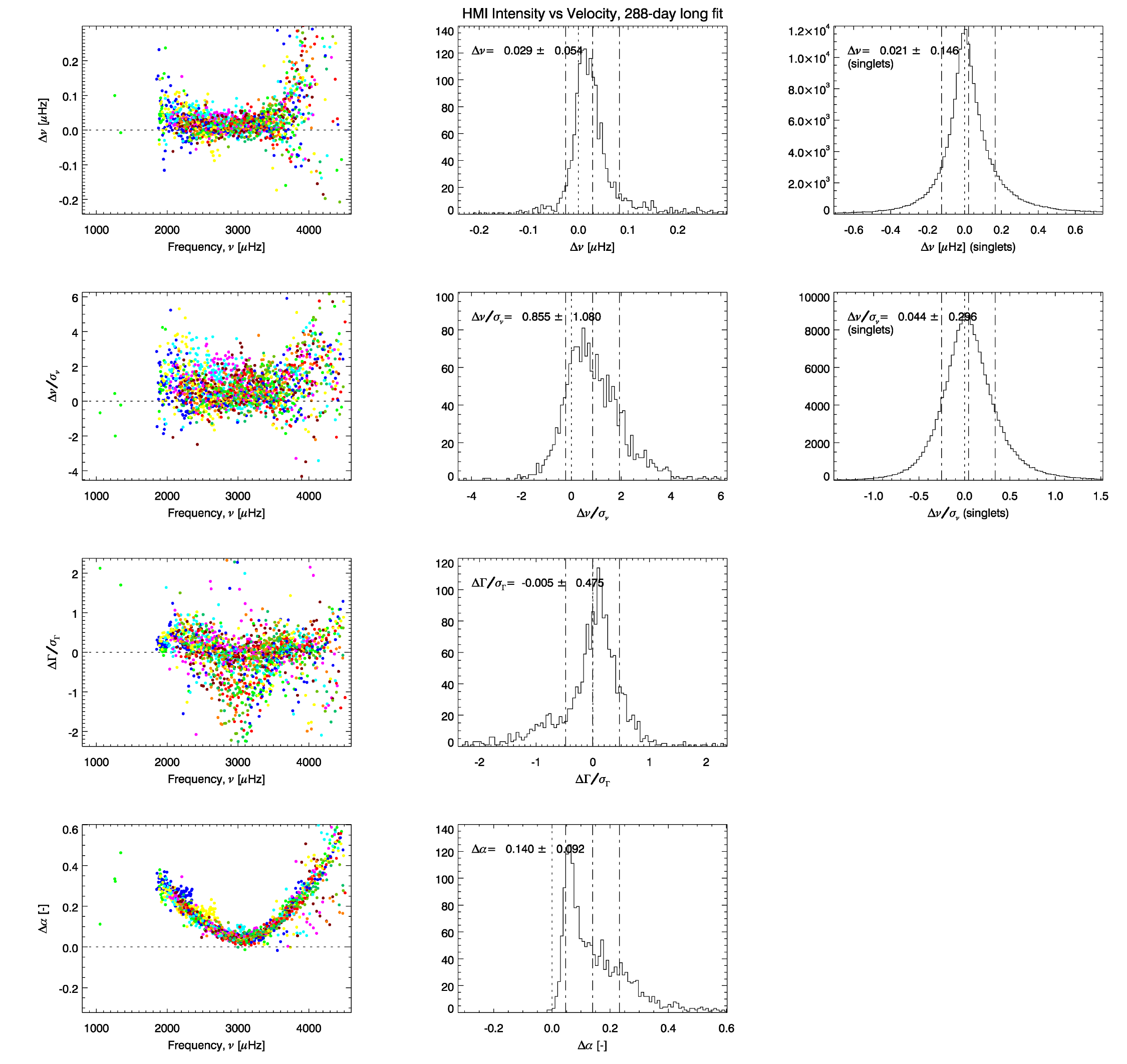}
\end{center}
\caption{Comparison of mode characteristics derived from fitting a 288-day
  long time series, as shown in Fig.~\ref{fig:compare-vel-int-72d} for the
  72-day long case, and also resulting from fitting co-eval intensity and
  velocity time series. 
  Similarly to the 72-day long case, the differences in asymmetries are large
  and show a clear and smooth dependence on frequency.
\label{fig:compare-vel-int-288d}}
\end{figure}

\Clearpage
\begin{figure}
\begin{center}
\includegraphics[width=.85\textwidth]{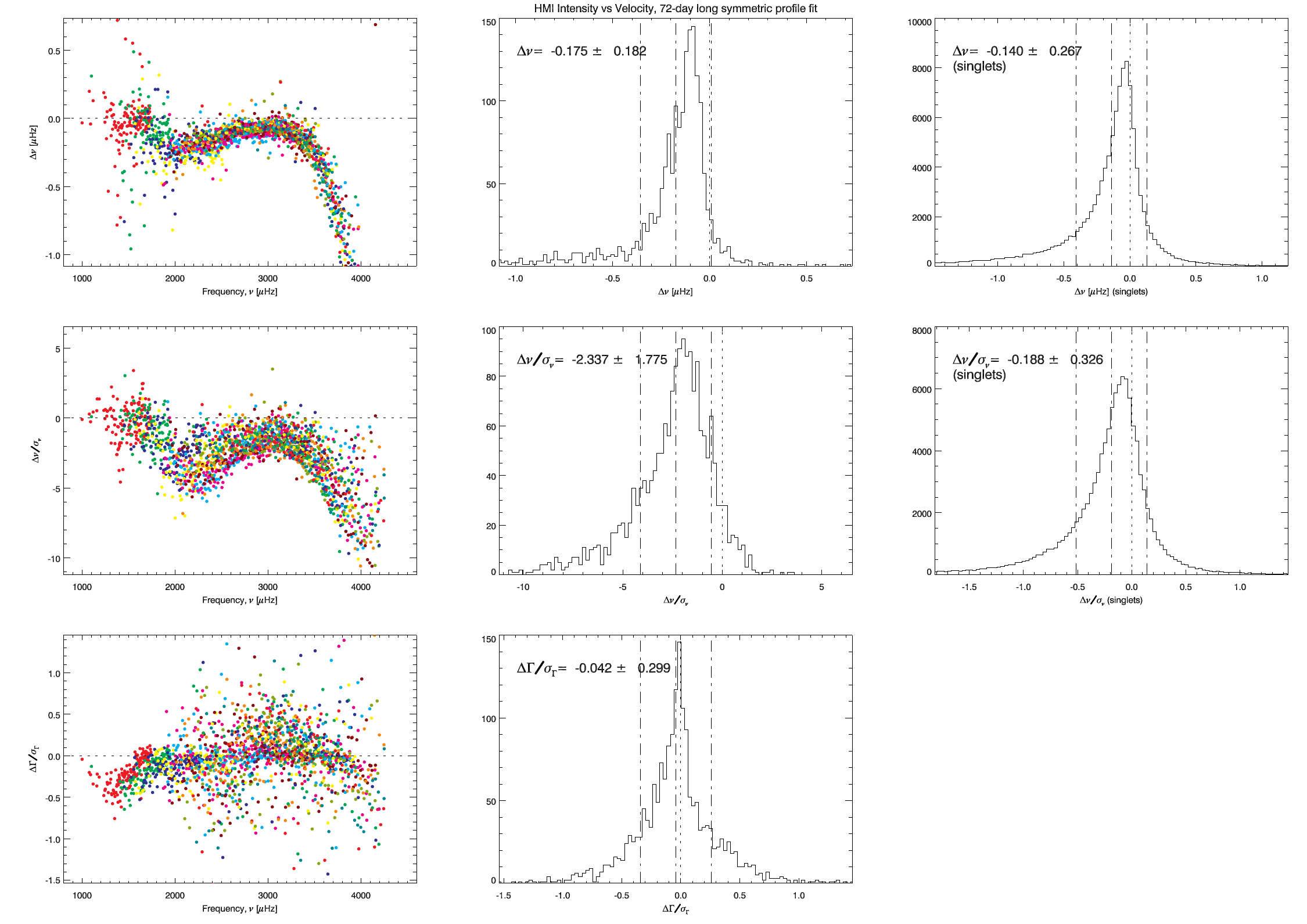}
\end{center}
\caption{Comparison of mode characteristics derived from fitting a 72-day long
  time series, as shown in Fig.~\ref{fig:compare-vel-int-72d}, but resulting
  from fitting co-eval intensity and velocity time series, using in both cases
  a symmetric peak profile ($\alpha_{n,\ell}=0$). The frequency differences
  become significant and systematic when ignoring the mode profile
  asymmetry. 
%
\label{fig:compare-vel-int-sym-72d}}
\end{figure}

\Clearpage
\begin{figure}
\begin{center}
\includegraphics[width=.85\textwidth]{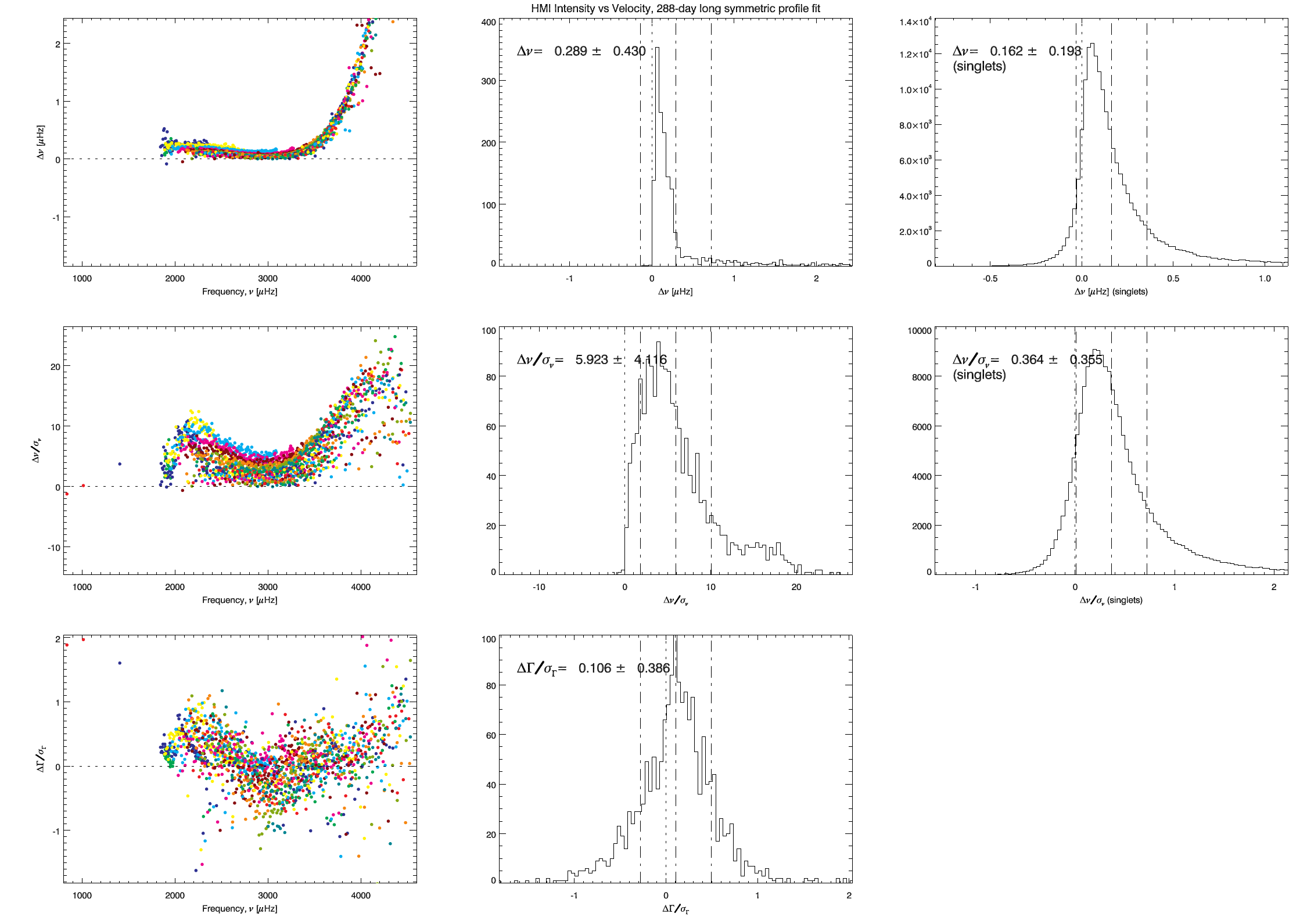}
\end{center}
\caption{Comparison of mode characteristics derived from fitting a 288-day
  long time series, as shown in Fig.~\ref{fig:compare-vel-int-sym-72d}, namely
  resulting from fitting co-eval intensity and velocity time series, using in
  both cases a symmetric peak profile ($\alpha_{n,\ell}=0$). The frequency
  differences are quite significant and systematic when ignoring the mode
  profile asymmetry.
%
\label{fig:compare-vel-int-sym-288d}}
\end{figure}

\Clearpage
\newcommand{\MTWO}[1]{\multicolumn{2}{c}{#1}}
\begin{table}
\begin{center}
\begin{tabular}{c|cc|cc}\hline
Length & Start Time       & End Time         & \MTWO{Duty Cycle (\%)} \\
 (day) & (TAI)            & (TAI)            & Velocity & Intensity \\\hline
    72 & 2010.04.30 00:00 & 2010.07.10 23:59 &  99.991  &  99.660   \\
       & 2010.07.11 00:00 & 2010.09.20 23:59 &  99.466  &  98.328   \\
       & 2010.09.21 00:00 & 2010.12.01 23:59 &  99.468  &  97.078   \\
       & 2010.12.02 00:00 & 2011.02.11 23:59 &  99.462  &  98.958   \\
\hline
   288 & 2010.07.11 00:00 & 2011.04.24 23:59 &  99.366  &  97.774   \\\hline
\end{tabular}
\end{center}
\caption{
{
Length, start and
  end time of fitted time series and their respective duty cycles. The duty
  cycles of the velocities time series correspond to the gap filled ones, the
  duty cycles of the intensity time series correspond to the raw series. 
}
\label{tab:fitranges}}
\end{table}






\newcommand{\MCX}[2]{\multicolumn{#1}{c|}{#2}}
\begin{table}
\begin{center}
{\footnotesize
\begin{tabular}{|l|r|r|r|r|}\hline\hline
Fitted profile     & \MCX{4}{Asymmetric} \\ \hline
Time series length & \MCX{4}{72 days}    \\
Data type/Leakage  & \MCX{4}{$V$/SU}     \\\cline{2-5}
Start time         & 2010.04.30 & 2010.07.11 & 2010.09.21 & 2010.12.02 \\
No.\ of singlets   &    205,530 &    206,251 &    204,956 &    204,969 \\
No.\ of multiplets &      2,297 &      2,296 &      2,287 &      2,294 \\\cline{1-5}
Data type/Leakage  & \MCX{4}{$I$/SU} \\\cline{2-5}
No.\ of singlets   &    149,420 &    148,457 &    146,072 &    147,952 \\
No.\ of multiplets &      1,679 &      1,669 &      1,649 &      1,675 \\\cline{2-5}
Data type/Leakage  & \MCX{4}{$I$/SGK} \\\cline{2-5}
No.\ of singlets   &    145,793 &    145,020 &    142,612 &    144,266 \\
No.\ of multiplets &      1,662 &      1,678 &      1,657 &      1,661 \\\cline{1-5}
\multicolumn{5}{c}{~}\\\cline{1-4}
%
Time series length & \MCX{3}{288 days}   \\
Start time         & \MCX{3}{2010.07.11} \\\cline{2-4}
Data type/Leakage  & \MCX{1}{$V$/SU} & \MCX{1}{$I$/SGK}  & \MCX{1}{$I_{\rm gf}$/SGK} \\\cline{2-4}
No.\ of singlets   & 281,977 & 202,420 & 202,719  \\
No.\ of multiplets &   2,386 &   1,682 &   1,682  \\\cline{1-4}
\multicolumn{5}{c}{~}\\\hline\hline
Fitted profile     & \MCX{4}{Symmetric} \\\hline
Time series length & \MCX{4}{72 days}   \\
Data type/Leakage & \MCX{4}{$V$/SU}     \\\cline{2-5}
Start time         & 2010.04.30 & 2010.07.11 & 2010.09.21 & 2010.12.02 \\
No.\ of singlets   &    206,227 &    206,265 &    203,374 &    204,858  \\
No.\ of multiplets &      2,287 &      2,285 &      2,278 &      2,281  \\
\hline
Data type          & \MCX{4}{$I$/SGK}   \\\cline{2-5}
No.\ of singlets   &  143,534  & 142,502  & 140,386  & 141,894 \\
No.\ of multiplets &    1,654  &   1,655  &   1,628  &   1,649 \\
\hline
\multicolumn{5}{c}{~}\\\cline{1-3}
Time series length & \MCX{2}{288 days}   \\
Start time         & \MCX{2}{2010.07.11} \\\cline{2-3}
Data type/Leakage  & \MCX{1}{$V$/SU} & \MCX{1}{$I$/SGK} \\\cline{2-3}
No.\ of singlets   & 282,787 & 196,639   \\
No.\ of multiplets &   2,389 &   1,670   \\\cline{1-3}
\end{tabular}
}
\end{center}
\caption{
{
Number of fitted singlets and derived multiplets for different fitting cases.
}
\label{tab:fitcstats}}
\end{table}

\Clearpage
\newcommand{\MX}[1]{\multicolumn{1}{c}{#1}}
\newcommand{\MZX}[1]{\multicolumn{1}{|c}{#1}}
\newcommand{\MXX}[1]{\multicolumn{1}{c|}{#1}}
\newcommand{\MTX}[1]{\multicolumn{2}{c|}{#1}}

\begin{table}
\begin{center}
{\footnotesize
\begin{tabular}{|r|rrrr|r@{~}l|}\hline\hline
Length & \MX{$\Delta \nu$} 
       & \MX{$\Delta \nu/\sigma_{\nu}$} 
       & \MX{$\Delta \Gamma/\sigma_{\Gamma}$}
       & \MXX{$\Delta \alpha/\sigma_{\alpha}$} & \MTX{Number of} \\
\MZX{[days]} & \MZX{[$\mu$Hz]} &  &  &  & \MTX{common modes} \\\hline
\multicolumn{5}{c}{~}\\*[-.95em]
\multicolumn{1}{c}{~} & \multicolumn{4}{l}{Asymmetric fitting, $I$, different
  leakage matrices, {\newtext i.e., SU vs SGK}} \\\hline
   72 &  0.000 $\pm$ 0.015 &    0.001 $\pm$ 0.017 & &                                       & 142,704 & singlets \\
      &  0.001 $\pm$ 0.014 &    0.001 $\pm$ 0.016 & &                                       & 141,645 & \\
      &  0.000 $\pm$ 0.015 &    0.001 $\pm$ 0.016 & &                                       & 139,598 & \\
      &  0.000 $\pm$ 0.015 &    0.000 $\pm$ 0.017 & &                                       & 141,164 & \\
  288 & -0.000 $\pm$ 0.008 &   -0.000 $\pm$ 0.015 & &                                       & 201,658 & \\\hline
   72 &  0.000 $\pm$ 0.029 &    0.004 $\pm$ 0.233 &   -0.023 $\pm$ 0.060 &   -0.001 $\pm$ 0.071 &   1,653 & multiplets \\
      &  0.001 $\pm$ 0.021 &    0.025 $\pm$ 0.231 &   -0.021 $\pm$ 0.058 &   -0.002 $\pm$ 0.061 &   1,651 & \\
      &  0.001 $\pm$ 0.023 &    0.014 $\pm$ 0.235 &   -0.017 $\pm$ 0.059 &   -0.004 $\pm$ 0.065 &   1,633 & \\
      &  0.002 $\pm$ 0.032 &    0.017 $\pm$ 0.220 &   -0.019 $\pm$ 0.060 &   -0.001 $\pm$ 0.065 &   1,647 & \\
  288 &  0.000 $\pm$ 0.006 &   -0.000 $\pm$ 0.113 &    0.017 $\pm$ 0.055 &    0.001 $\pm$ 0.043 &   1,679 & \\\hline

\multicolumn{5}{c}{~}\\*[-.95em]
\multicolumn{1}{c}{~} & \multicolumn{4}{l}{Asymmetric fitting, $I$, 
  {\newtext gap filled vs no gap filling}} \\\hline
  288 &  0.001 $\pm$ 0.013 &    0.001 $\pm$ 0.028 & &                                            & 200,909 & singlets \\\hline
  288 &  0.001 $\pm$ 0.006 &    0.023 $\pm$ 0.125 &    0.005 $\pm$ 0.035   &   0.005 $\pm$ 0.045 &   1,676 & multiplets \\\hline

\multicolumn{5}{c}{~}\\*[-.95em]
\multicolumn{1}{c}{~} & \multicolumn{4}{l}{Asymmetric fitting, $I-V$} \\\hline

   72 &  0.015 $\pm$ 0.236 &    0.014 $\pm$ 0.289 & &                  &   115,347 & singlets \\
      &  0.015 $\pm$ 0.239 &    0.011 $\pm$ 0.298 & &                  &   115,299 & \\
      &  0.014 $\pm$ 0.238 &    0.011 $\pm$ 0.294 & &                  &   114,003 & \\
      &  0.012 $\pm$ 0.236 &    0.008 $\pm$ 0.292 & &                  &   114,621 & \\
  288 &  0.021 $\pm$ 0.146 &    0.044 $\pm$ 0.296 & &                  &   191,979 & \\\hline
   72 &  0.044 $\pm$ 0.113 &    0.470 $\pm$ 0.893 &   -0.065 $\pm$ 0.401 & &   1,637 & multiplets \\
      &  0.049 $\pm$ 0.152 &    0.455 $\pm$ 0.854 &   -0.021 $\pm$ 0.354 & &   1,658 & \\
      &  0.035 $\pm$ 0.096 &    0.434 $\pm$ 0.864 &   -0.033 $\pm$ 0.370 & &   1,637 & \\
      &  0.033 $\pm$ 0.097 &    0.384 $\pm$ 0.844 &   -0.040 $\pm$ 0.369 & &   1,640 & \\
  288 &  0.028 $\pm$ 0.053 &    0.856 $\pm$ 1.075 &   -0.008 $\pm$ 0.482 & &   1,674 & \\\hline

\multicolumn{5}{c}{~}\\*[-.95em]
\multicolumn{1}{c}{~} & \multicolumn{4}{l}{Symmetric fitting, $I-V$} \\\hline

   72 &  0.118 $\pm$ 0.262 &    0.144 $\pm$ 0.312 & &                    &   114,581 & singlets \\
      &  0.128 $\pm$ 0.265 &    0.153 $\pm$ 0.318 & &                    &   114,179 & \\
      &  0.124 $\pm$ 0.263 &    0.149 $\pm$ 0.315 & &                    &   112,974 & \\
      &  0.119 $\pm$ 0.261 &    0.146 $\pm$ 0.312 & &                    &   113,707 & \\
  288 &  0.161 $\pm$ 0.193 &    0.343 $\pm$ 0.350 & &                    &   188,372 & \\\hline
   72 &  0.183 $\pm$ 0.225 &    2.133 $\pm$ 1.724 &   -0.040 $\pm$ 0.416 & &     1,643 & multiplets \\
      &  0.173 $\pm$ 0.175 &    2.233 $\pm$ 1.729 &   -0.004 $\pm$ 0.346 & &     1,642 & \\
      &  0.165 $\pm$ 0.178 &    2.198 $\pm$ 1.780 &   -0.016 $\pm$ 0.375 & &     1,621 & \\
      &  0.183 $\pm$ 0.226 &    2.120 $\pm$ 1.726 &   -0.019 $\pm$ 0.379 & &     1,636 & \\
  288 &  0.203 $\pm$ 0.256 &    5.395 $\pm$ 4.058 &    0.076 $\pm$ 0.483 & &     1,662 & \\\hline\hline
\end{tabular}
}
\end{center}
\caption{
Mean and standard deviation around the mean of mode characteristics' raw or
scaled differences, computed using singlets or multiplets values, whether (i)
using different leakage matrix evaluations, (ii) using gap filling or not,
(iii) using intensity or co-eval velocity observations and fitting asymmetric
profiles, and (iv) again using intensity or co-eval velocity observations but
fitting symmetric profiles.
 \label{tab:diffs}}
\end{table}

\end{document}